\documentstyle[11pt,aaspp4,epsf]{article}

\def\erg{\ {\rm erg}}
\def\cm{\ {\rm cm}}
\def\km{\ {\rm km}}

\def\s{\ {\rm s}}
\def\hst{{\it HST}}
\def\etal{{\it et al.}}
\def\Msun{M_{\sun}}
\def\angstrom{\ {\rm \AA}}
\def\subr #1{_{{\rm #1}}}
\def\supr #1{^{{\rm #1}}}
\def\psubm{p_{\rm m}}
\def\dpmdr{d\psubm/dr}

\def\secspt{\farcs}
\def\minspt{\farcm}
\def\spt{\fs}

\def\simless{\lesssim}

\font\bbital=cmbxti10 scaled\magstep1

\slugcomment{To appear in the {\it Astronomical Journal}, April 1997}

\begin{document}

\title{Deep {\bbital HST}/FOC Imaging of the Central Density Cusp\nl
of the Globular Cluster M15\footnote{Based on observations with the
NASA/ESA {\it Hubble Space Telescope}, obtained at the Space Telescope
Science Institute, which is operated by AURA, Inc., under NASA
contract NAS 5-26555.}}

\author{Craig Sosin and Ivan R.\ King}
\affil{Astronomy Department, University of California, Berkeley, CA 94720}
\affil{Electronic mail: csosin@astro.berkeley.edu, king@astro.berkeley.edu}

\begin{abstract}
Using the Faint Object Camera on the repaired {\it Hubble Space
Telescope,} we have observed two fields in the globular cluster M15:\
the central density cusp, and a field at $r = 20''$.  These are the
highest-resolution images ever taken of this cluster's dense core, and
are the first to probe the distribution of stars well below the
main-sequence turnoff.  After correction for incompleteness, we
measure a logarithmic cusp slope ($d \log \sigma / d \log r$) of
$-0.70 \pm 0.05$ (1-sigma) for turnoff ($\sim0.8 \Msun$) stars over
the radial range from $0\secspt3$ to $10''$; this slope is consistent
with previous measurements.  We also set an approximate upper limit of
$\sim1$\secspt5 (90\% confidence limit) on the size of any possible
constant-surface-density core, but discuss uncertainties in this limit
that arise from crowding corrections.  We find that fainter stars in
the cusp also have power-law density profiles:\ a mass group near $0.7
\Msun$ has a logarithmic slope of $-0.56 \pm 0.05$ (1-sigma) over the
radial range from $2''$ to $10''$.  Taken together, the two slopes are
not well matched by the simplest core-collapse or black-hole models.
We also measure a mass function at $r = 20''$, outside of the central
cusp.  Both of the FOC fields show substantial mass segregation, when
compared with a mass function measured with the WFPC2 at $r = 5'$.  In
comparing the overall mass functions of the two FOC fields and the
$r=5'$ field, we find that the radial variation of the mass function
is somewhat less than that predicted by a King--Michie model of the
cluster, but greater than that predicted by a Fokker--Planck model
taken from the literature.
\end{abstract}


\newpage

\section{Introduction}

For years, studies of M15 have guided our understanding of the
dynamical evolution of globular clusters.  Observations have shown
that the cluster's stellar density rises in its center in a power-law
cusp, with no sign of a ``flat,'' constant-surface-density core at the
resolution attainable from the ground (Djorgovski \&
King\markcite{djk84} 1984, Lugger\markcite{lug87} \etal\ 1987).  Most
researchers have considered M15 to be in a state of core collapse, in
which the cluster core's negative heat capacity has resulted in its
runaway contraction.  The collapse would eventually be halted by
energy input from binary stars (see the references in Hut \&
Makino\markcite{iau} 1996).  Alternatively, M15 could harbor a massive
($\sim10^3 \Msun$) black hole in its center; the cusp would then be a
result of the black hole's deep potential well (Bahcall \&
Wolf\markcite{bw76} 1976).

The increase in resolving power provided by the {\it Hubble Space
Telescope} ({\it HST}) can help to determine the nature of the density
cusp.  The principal goal of \hst\ observations of M15 has been to
resolve stars near the main-sequence turnoff, since the large number
of such stars should trace the stellar density profile much more
clearly than the handful of bright giants seen from the ground.

Early \hst\ results, though, provided more confusion than clarity:\
until the 1993 repair, most of the light of the bright stars was
spread into an extended ``halo'' that obscured the view of the faint
stars.  Lauer \etal\ (1991)\markcite{lau91} attempted to subtract
these bright stars from a $U$-band Planetary Camera image of the M15
cusp, and claimed the detection of a ``flat'' (constant
surface-density) core of radius 2\secspt2 in the residual light.
Yanny\markcite{yan94} \etal\ (1994), however, found a serious flaw in
this procedure:\ stellar-photometry software attributes too much of
the light to the bright stars in a crowded field, at the expense of
their faint neighbors.  The bright stars are thus oversubtracted in
the densest region near the cluster center, creating the illusion of a
flat core.

Guhathakurta\markcite{GYSB} \etal\ (1996, hereafter GYSB) followed up
the Yanny \etal\ study by taking post-repair WFPC2 images of the M15
cusp, and found that the stellar density rises all the way into $r
\simeq$ 0\secspt3.  Moreover, GYSB found that the cusp surface density
is well fit by a power law of slope $-0.82 \pm 0.12$---quite close to
the slope of $-0.75$ expected for stars near a massive black hole, but
also consistent with the outcome of core collapse in the presence of
dark objects of mass $\geq 1 \Msun$ (Cohn\markcite{hc85} 1985,
Grabhorn \etal\ \markcite{gra92}1992).

Our main goal in this paper is to present an analysis of a set of
images of the M15 cusp, taken with the Faint Object Camera (FOC)
aboard the repaired \hst.  The FOC samples the \hst\ point-spread
function (PSF) much better than the WFPC2:\ its pixel scale is
$0\secspt014$/pixel, compared to the Planetary Camera chip's
$0\secspt044$/pixel (the FWHM of the $V$-band PSF is $0\secspt04$).
In the central $2''$ of M15, the typical separation of stars with $19
< V < 21$ is $\sim 0\secspt15$, so the FOC can separate many more
close stellar pairs.

We begin by describing the images themselves.  We then describe our
procedure for extracting as much photometric information as possible
from the images, and the extensive artificial-star experiments that
accompany this procedure.  We then use these data to construct the
surface-density profiles and mass functions of M15.  Finally, we
construct a simple dynamical model of M15, and compare our results
with a Fokker--Planck model from the literature.

\section{Observations}

M15 was observed by \hst\ on 27 September 1994, with the COSTAR
optical correction system in place.  (For details of the design and
performance of the camera and correction system, see
Greenfield\markcite{gre91} \etal\ 1991 and
Jedrzejewski\markcite{jed94} \etal\ 1994.)  Three $\sim 7 \times
7$-arcsecond FOC images were taken through each of the F430W (FOC
``$B$'') and F480LP (FOC ``$V$'') filters; the exposure times ranged
from 2020.75 to 2519.75 seconds.  All of the images were geometrically
corrected and flatfielded by the STScI Routine Science Data Processing
(RSDP) pipeline.

The inner two FOC fields overlap slightly, and together cover a $\sim
13 \times 7$-arcsec region around the cluster center.  Their combined
$V$ images are shown in Figure \ref{M15Inner}.  We reduced these
images in their original form, but then immediately transformed the
results to a common coordinate system; the combined field will be
called the ``central field'' in what follows.  In this study, we
analyzed only the $V$ images of the central field, since the $B$
images suffer badly from detector saturation.

The third $B$ and $V$ pair of images were taken of a field 20
arcseconds out from the cluster center; the $V$ image is shown in
Figure \ref{M15Outer}.  This field is devoid of bright stars, so that
both the $B$ and $V$ images are suitable for analysis.  The lesser
degree of stellar crowding and the lower background of scattered light
allow the photometry of the $r=20''$ field to go to $V = 24,$ one or
two magnitudes deeper than in the central field.

A closeup of the central $\sim 3.5 \times 3.5$ arcseconds of M15 is
shown in Figure \ref{CenterCloseup}, with stars of various $V$
magnitudes indicated.  Note that the large crescent-shaped ``objects''
are bright, saturated stars; their shape is a result of peculiarities
of the FOC photon-event detection logic.  The FOC saturates at high
count {\it rate}, rather than at high {\it numbers} of counts, so the
saturation cannot be removed by taking an additional set of short
exposures.  (FOC images can be taken through neutral-density filters,
if information about bright stars is needed.  In the case of M15,
short exposures of the center have already been taken with the WFPC2,
by GYSB.)

Looking at Fig.\ 3, one's immediate impression is that the stellar
density continues to rise into the very center of the cluster, and
does not level off at $r \simeq 2'',$ as suggested by pre-repair work
(Lauer\markcite{lau91} \etal\ 1991).  To some degree, this impression
is a result of the higher background level created by scattered light
from a few bright stars, whose chance placements near the center do
not imply a significantly higher stellar density.  One goal of our
analysis will be to use the larger number of turnoff-magnitude stars
to determine whether a $2''$ core is present.
 
\section{Photometry}

In this section and the next, we describe our procedure for image
reduction and analysis.  We begin, in \S 3, by describing the
photometry---in particular, methods for finding faint stars and
rejecting false detections in crowded regions.  In \S 4, we describe
the artificial-star experiments that allow us to correct for the
errors induced by crowding.

The crowding of stars in this field has led to some controversy over
observational results in recent years.  With this fact in mind, we
devoted a great deal of attention to analysis methods, and will
describe them in some detail.  Readers interested only in the results
of the analysis will find those results in the latter half of \S 5 and
in \S 6.

\subsection{The central field}

First, it is apparent from Figures \ref{M15Inner} and \ref{M15Outer}
that parts of the images ($\sim10\%$) are unusable because of the
presence of saturated ($V < 18$) stars.  We identify these regions by
hand before beginning, and exclude them from any further analysis.

The stellar photometry package DAOPHOT (Stetson\markcite{pbs87} 1987,
\markcite{pbs92}1992) has become a standard tool for globular-cluster
image analysis, and we adopt it here.  DAOPHOT was originally designed
for ground-based images, in which the point-spread function (PSF) is
nearly Gaussian in its center, and decreases monotonically at larger
radii.  \hst\ is diffraction-limited, though, and the FOC has a small
enough pixel size that the innermost diffraction rings are
``resolved'' quite clearly (see Fig.\ \ref{CenterCloseup}).  When used
with FOC images, the FIND algorithm in DAOPHOT often identifies
several false ``objects'' on the first diffraction ring of each star,
and thus finds ``clusters'' of objects four or five pixels apart.

It is important, then, to reject such false detections, while throwing
away as few {\it real} objects as possible (since many objects do
happen to fall on the diffraction ring of another star).  We have
developed a three-pass procedure to do so:

{\bf First pass.}  To begin, we use DAOPHOT FIND to identify stars,
and also ``find'' nearly three times as many false detections.  We
then compare the central pixel of each detection to the highest pixel
in a annulus between 4.5 and 5.5 pixels from that central pixel.
False detections on diffraction rings will always have a high
peak---the star's true PSF core---in the annulus, while most true
detections will not (except for those that have bright close
neighbors).  We exclude, for the remainder of the first pass, any
detection whose highest pixel in the annulus is more than twice as
high as its central pixel.  (Isolated real stars, on the other hand,
have a central pixel that is 5--10 times higher than the highest pixel
in the annulus.)

The remaining list contains perhaps 80\% of the stars that will
ultimately be identified, and has {\it no} false detections.  We then
complete the standard DAOPHOT reduction sequence using this
preliminary list of objects:\ aperture photometry, determination of
the PSF using isolated stars (selected manually), and PSF-fitting, to
obtain a preliminary magnitude for each star.  (The PSF determination
was actually iterated two or three times, with neighbors of the PSF
stars subtracted on the later iterations.)

Figure 4a shows a small section of the original image of our central
field.  Figure 4b shows the same section after stars have been
subtracted using the first-pass magnitudes.  Note that a number of
faint stars have been missed, at this point.  The circle in Fig.\ 4a
indicates a star with another star ``hidden'' in its diffraction ring;
this ``hidden'' star is mistaken for a diffraction-ring feature by the
first pass.

{\bf Second pass.}  We then use the results of the first pass to
subtract a scaled copy of the PSF from each star in the preliminary
list from the previous pass---but we subtract only those pixels that
lie $\ge$ 4 pixels away from the central pixel of the PSF.  This has
the effect of removing the diffraction rings from the image (the first
ring is at a radius of 5 pixels), while leaving the central peaks.
Faint stars hidden {\it in} the diffraction rings of a brighter
neighbor also remain.  A section of this ``ring-subtracted'' image is
shown in Figure 4c (note that the ``hidden'' star from Fig.\ 4a shows
up clearly on Fig.\ 4c).  We then run FIND on this ring-subtracted
image, producing a second preliminary list of stellar positions.

We then use the second preliminary star list as input to the usual
DAOPHOT sequence---aperture photometry and PSF-fitting---for a second
time.

(Unfortunately, the brightest non-saturated stars are slightly
nonlinear in their centers, so that their rings do not subtract
cleanly, and false ``objects'' are found in the poorly-subtracted
rings on the second pass.  We therefore ignore all found neighbors of
such stars from here on---removing a further few percent of the
central field from consideration.  All stars with $V$ magnitudes
brighter than 18.2 [measured in the second-pass aperture photometry
step] were considered to be nonlinear; we chose this limit from a
visual inspection of their appearance on the ring-subtracted image.)

{\bf Third pass.}  The PSF-fitting step in the second pass produces an
image with all stars found thus far subtracted; see Figure 4d for an
example.  (The ``hidden'' star in Fig.\ 4a has now been successfully
fit and subtracted.)  An inspection of this image reveals a number of
``double-lobed'' residuals, where stars separated by $\simless 3$
pixels were reduced as a single star by the previous pass (one such
blend is indicated by a circle in Fig.\ 4d).

We identify these blended close pairs on the second-pass subtracted
image (Fig.\ 4d) as follows:\ At each position where a star was
subtracted, our software first determines the angle $\phi$ along which
the residual light is most elongated.  Then, it measures (1) the
number of residual counts $F\subr{a}$ in a zone within $15^{\circ}$ of
this major axis, out to a radius of 5 pixels, (2) the number of counts
$F\subr{b}$ within $15^{\circ}$ of the minor axis within the same
radius, and (3) the mean distance $\bar r$ of the residual flux from
the stellar center.

Locations with large $F\subr{a} - F\subr{b}$ are either blended pairs
mistakenly reduced as a single star, or single objects that subtracted
poorly for some other reason (most often, because they are nonlinear
in their centers).  The quantity $\bar r$ distinguishes between these
possibilities: the residual light from blends is at large radius,
while that from a poorly fit single star is near the star's center.
We thus identify any subtracted ``star'' with $F\subr{a} - F\subr{b} >
800$ counts and $\bar r > 3$ pixels as a candidate blend.  (A better
algorithm might be to normalize the $F$ difference by the flux of the
``star,'' but in practice, the 800-count criterion had the desired
effect of finding the most obvious blends, without finding noise.)

We then split each candidate blend into two stars in our list,
separating the two by one FWHM of the PSF, at position angle $\phi$.
The resulting list is run through a third and final stage of
PSF-fitting, to refine the positions and magnitudes of the close
pairs, and it is the output list from this step that we take as the
final result of our image reduction.  A portion of the final
subtracted image is shown in Figure 4e.  Note that some of the blends
have been removed (such as the blend indicated in Fig.\ 4d), but not
all.  Artificial-star experiments will allow us to correct the star
counts for the few that have been missed.

The reduction scheme may seem tedious, but it is not very ``long'' in
terms of CPU time (a complete run, excluding the initial interactive
measurement of the PSF, takes about 15 to 20 minutes on our
workstation).  Note that since every step is entirely automated (after
the PSF measurement), artificial-star experiments require no
additional human effort.

\subsection{The 20$''$ field}

We also performed stellar photometry on the $B$ and $V$ images of the
$r=20''$ field.  The reduction scheme was the same, except that the
entire third pass---the splitting of close pairs, separated by less
than 3 pixels---was skipped.  (Since the stars on this field are much
less crowded, there are far fewer close blends, and searching for them
would have been a waste of time.  Our artificial-star experiments
later confirmed this impression, and allowed us to correct for the few
close pairs that were present.)

We then calibrated the photometry (of both the central and $r=20''$
fields).  The FOC has been calibrated by medium- and narrowband
observations of spectrophotometric standards (see
Greenfield\markcite{gre91} \etal\ 1991).  The resulting filter
sensitivities are written into each image header by the RSDP pipeline;
those photometric header keywords allowed our stellar magnitudes to be
adjusted to the \hst\ instrumental system.  An \hst\ magnitude
$m_\lambda$ is defined as $m_\lambda = -2.5 \log f_\lambda - 21.1,$
where $f_\lambda$ is expressed in $\erg \cm^{-2} \s^{-1}
\angstrom^{-1}$ at the filter's ``pivot wavelength'' ($4111 \angstrom$
for F430W, and $5254 \angstrom$ for F480LP).

We then converted our magnitudes from the \hst\ instrumental system to
Johnson $B$ and $V$, using a transformation calculated by Jay Anderson
(see King, \markcite{irk94}Anderson, \& Sosin 1994).  The resulting
color--magnitude diagram of the $r=20''$ field is shown in Figure
\ref{OuterCMD}, including a small correction to be described below.
Note that the FOC $V_{480}$ magnitude is almost identical to Johnson
$V$, and we use $V_{480}$ and Johnson $V$ interchangeably in what
follows.

Durrell \& Harris (1993, \markcite{DH93}hereafter DH93) have presented
a deep ground-based CMD of a field $7'$ from the center of M15.  Their
``smoothed'' fiducial main-sequence line is shown in Figure 5 (dashed
line), along with the ([Fe/H] $= -2.26,$ $t = 15 $ Gyr) Bergbusch \&
Vandenberg (1992, \markcite{BV92}hereafter BV92) isochrone (solid
line).  DH93 found this isochrone to fit the sequences in their CMD,
with $(m - M)_V = 15.40,$ $E(B-V) = 0.10,$ and a small ``color shift''
of 0.015.

We needed to shift our own stellar colors redward by 0.12 mag to match
the DH93 fiducial; this shift is comparable to the FOC calibration
uncertainty.  With this shift, our objects fit the isochrone well, so
we adopt the same distance and reddening as DH93, and later will use
the BV92 isochrone to convert luminosities to masses.  Note that our
sequence deviates to the red from the DH93 sequence at the bright end,
because of the onset of detector nonlinearity in $B$; the $B$ images
of these stars do in fact appear slightly distorted.  The $V$
magnitudes of the stars plotted are unaffected by nonlinearity.

Note the presence of an anomalously blue object at $V =$ 20.2, $B-V =$
0.3 in Figure 5; it appears at (232, 102) on the $r = 20''$ $V$ image,
is indicated by a circle in Fig.\ 2, and has J2000 coordinates
$(21\supr{h}29\supr{m}58\spt27, 12^{\circ}10'19\secspt9)$.  Its $B$
magnitude is strongly affected by nonlinearity, so it is probably even
bluer than shown on Fig.\ 5.  This object could be similar to the blue
objects found in the core of M15 by\markcite{dmp96} De Marchi \&
Paresce (1996).

Tables of positions and magnitudes of stars in both fields are
available from the first author.

\section{Corrections for crowding}

No reduction scheme can detect every star in a field.  For the star
counts produced by the procedure described in \S 3 to be meaningful,
they must be corrected for the fraction of stars missed entirely, and
for the fraction that are measured so poorly that they jump in or out
of the magnitude range under consideration.  Alternatively, the
observed counts could be compared directly with a model of the density
distribution in which these crowding effects had been included.  Since
we will use both strategies in \S 5 and \S 6, we first discuss the
methods of correction here.

\subsection{Artificial-star experiments}

Artificial-star experiments, in which the images are re-reduced many
times with a few artificial stars of known magnitude and position
added each time, have become a standard technique for quantifying the
effects of crowding in observations of star fields (e.g.,
incompleteness).  Since our images were reduced entirely by software,
performing many such experiments required minimal effort.

In each experiment, we added $\sim150$ stars of random magnitude and
position to each image, and re-reduced from scratch.  (We ensured that
the artificial stars added in a single experiment did not interfere
with one another, by reselecting any position that was within 32
pixels of an artificial star that had already been added.)  In the
$r=20''$ field, artificial stars were added at the same position in
both $B$ and $V$ frames, with a color that would put the star on the
main-sequence ridgeline.  No color was needed for stars in the central
field, since we did not use its $B$ image.

After the images with artificial stars had been reduced, we recorded
whether the artificial stars were recovered, and if so, their
positions and magnitudes.  We also recorded the positions and
magnitudes of any {\it real} neighbor of the artificial stars.  This
latter step allowed us to investigate the blending of stellar images:
if an artificial star happened to fall within 3 pixels of a real star,
the two were sometimes identified as a single object by the reduction
procedure.  In such cases, we considered the object in the output list
to be the {\it brighter} of the two.  In other words, if the
artificial star was brighter than its real neighbor, then the
artificial star was recorded as having been recovered, with the
magnitude of the blended object in the output list.  On the other
hand, if the artificial star was the fainter of the two, it was
considered to have been lost.  This scheme eliminates false large
jumps in magnitude arising from misidentifications, and ensures that
correction factors will be derived correctly by the matrix method
described in \S 4.3.

In the central field, we wish to measure star counts (and, therefore,
completeness statistics) as a function of both magnitude and position,
which required us to analyze a large number of artificial stars
($\sim6\times10^4$).  For the $r=20''$ image, we will obtain only one
LF for the entire field, so a smaller number sufficed ($\sim10^4$).

\subsection{Causes of incompleteness}

We may use the results of our artificial-star experiments in two ways.
First, we could correct our {\it observed}, ``raw'' star counts for
incompleteness, to obtain the {\it corrected} number of stars actually
present.  Alternatively, we could begin with a {\it model} of the true
number of stars present (usually as a function of position or
magnitude), and then test the model by using the artificial-star
results to predict how many of those stars would be detected by our
reduction procedure, and then comparing that prediction with the
observed counts.

While the first approach is both more common and more straightforward,
the statistical properties of the corrected counts that it produces
make those counts difficult to use (\S 5.4).  The second approach
avoids these problems, but it requires an understanding of the reasons
that some stars were detected while others were not, for adequate
predictions to be made.  Let us consider the two approaches.

First, correcting observed counts for incompleteness is quite simple:
we use the artificial-star results to find a completeness fraction
$c$, defined as the ratio of the number of artificial stars recovered
in a given magnitude range to the number originally added in the same
range, and then correct the observed counts using $c$.  For example,
if we count 1000 stars in a region, and find a completeness fraction
of 0.8, then our estimate for the true number of stars is $ 1000 / 0.8
= 1250$.  The procedure can be generalized to handle multiple
magnitude ranges, when measuring a luminosity function (\S4.3).

On the other hand, suppose that we wish to use the other approach:
that we have a cluster model that predicts the number of stars in some
region, and that we want to use the artificial-star results to further
predict how many of those stars we would actually detect, if we
observed a cluster described by that model.  When we use this approach
in \S5.5, the models will be density profiles (star counts as a
function of radius), but for simplicity, let us consider here a model
that consists of a given number of stars uniformly distributed over a
region of the sky.  Our goal, then, is to see whether the number of
detectable stars is {\it incompatible} with our observed number of
stars; if so, we can rule out the model in question.

For example, we might be interested in whether the region mentioned
above could contain 1100 stars, rather than 1250.  The region might be
the central $2''$ of a globular cluster, and the difference between
1100 and 1250 stars might be the difference between the numbers of
stars in a $2''$ core and in a cusp.  In this case, we would want to
know whether we could rule out the $2''$ core.

Now, we cannot turn our model's total number of stars, 1100, into a
prediction of the number we would detect, without an understanding of
the {\it causes} of the incompleteness: the circumstances under which
a particular star would or would not be found.  To show this, let us
look at two extreme cases, before discussing our actual M15 data.

In the first case, imagine that a few brighter stars appear in the
real image, blocking our view of 20\% of the (fainter) stars of
interest (giving us $c=0.8$).  (Call this situation Case I.)  In this
case, we {\it know} those brighter stars' positions, and thus know
that they will block the {\it same} fraction of the fainter stars, no
matter how many of the latter happen to be there.  We can conclude
that if 1100 fainter stars were present (as in our alternative model),
then we would measure $1100 \times 0.8 = 880$ of them---a number quite
different from our measurement of 1000 in the real image.  If the
difference were large enough, we might rule out the 1100-star model.

As a second case, though, imagine that the cause of the incompleteness
was crowding of the faint stars themselves.  In other words, suppose
that 20\% of the faint stars in the real 1250-star image happened to
fall along the same line-of-sight as other faint stars, so that the
resulting high fraction of blends lowered $c$ to 0.8.

In this case---Case II---it is not clear what the completeness
fraction would be if only 1100 stars were present, since the frequency
of blends depends upon the stellar density.  We might be tempted to
apply the $c$ of 0.8 to the 1100-star model, as in the previous case,
but this would not be correct.  Since an 1100-star image would be less
crowded than the 1250-star image we have already analyzed, the
appropriate $c$ in the 1100-star case would be somewhat higher than
0.8.  Unfortunately, {\it there is no way to know how much higher,}
without running a new set of simulations of the reduction process with
1100-star artificial images.

Leaving aside the details of how one might do such simulations,
suppose that they told us that the completeness fraction at the lower
density was, say, 0.9.  Our prediction would then be that we would
detect $(1100)(0.9) = 990$ stars---almost exactly the number that we
{\it did} detect in our actual image (1000).

So, is it possible to say which model fits better, in a Case II field:
1250 stars with 80\% completeness, or 1100 stars with 90\%
completeness?  While the predicted number of counts is nearly the
same, the measurement of $c$ {\it does} distinguish the two
(specifically, the measured $c$ of 0.8 is much lower than the
1100-star model's 0.9).

The difficulty in rejecting the 1100-star model, though, is twofold.
First, more work was required (an additional set of simulations).
Second, the {\it significance} of the disagreement between the
observed quantities and the ill-fitting model's prediction is
difficult to determine, because the uncertainty of the completeness
fraction is not well known.  There could easily be systematic
differences between the real images and the simulations, since the
latter would have to be created from scratch.  (Artificial-star
experiments, on the other hand, begin with the real image, so such
differences would be much smaller.)  In particular, the FOC's
nonlinearity and saturation would be very difficult to simulate
realistically.

To sum up: Artificial-star experiments allow us to correct observed
counts for incompleteness in {\it both} cases.  They also allow us to
incorporate incompleteness into a model, {\it if and only if} the
incompleteness is similar to Case I (i.e., dominated by bright stars).
In a region where blending of faint stars is important,
artificial-star results cannot be applied to alternative models;
incompleteness in those models most be simulated in some other way.

Before proceeding further, then, it is appropriate to determine which
case applies to our M15 images, especially the innermost few
arcseconds.  Near the cluster center, the image is nearly covered by
faint stars (see Fig.\ \ref{CenterCloseup}, and recall that the areas
covered by saturated stars were excluded from analysis entirely), so
one might wonder whether the region was more like Case II than Case I.
However, we will show later that the region is actually closer to Case
I than II: the incompleteness of the stellar sample we will use for
profile-fitting is dominated by the effect of stars brighter than that
sample (but not so bright as to be saturated).  We will therefore go
on to describe our methods for using the artificial-star results in
more detail---knowing in advance that those results will not need to
be replaced by an impractically large set of simulations when we fit
models in \S5.

\subsection{Recovering the true LF from the data}

To correct the raw star counts for crowding---in a more general case
than the simple example given in the last section---we can proceed as
follows: Let the true luminosity function (LF) of stars in an observed
field be expressed as a vector $T$, where $T_j$ is the number of stars
observed in magnitude bin $j$.  Let vector $R$ be the ``raw'' LF: the
number of stars that actually appear in each bin of our photometry
list.  Two effects cause $R$ to differ from $T$.

The first is scatter in the photometry: photon statistics and crowding
cause stars to be measured brighter or fainter than their true
magnitudes.  In this case, features in the true LF $T$ will be blurred
in the observed LF $R$.  This effect is sometimes called
``bin-jumping''.

The matrix method of\markcite{dru88} Drukier \etal\ (1988) allows us
to correct for bin-jumping, and recover $T$ from $R$.  We use the
output of our artificial-star experiments to construct a transition
matrix $A$, where element $A_{jk}$ measures the probability that a
star whose true magnitude is in bin $j$ will be measured to be in bin
$k$.  Once we have $A$, its inverse allows us to obtain $T$ from $R$
(i.e., since $R = A \cdot T$, $T = A^{-1} \cdot R$).  Note that we
need not ``guess'' any particular LF before running artificial-star
experiments, since the absolute number of artificial stars in each bin
never enters the calculation.

The second reason that $R$ will differ from $T$ is that some stars are
entirely undetected by the reduction procedure---some blended with
neighbors, and others lost in the noise.  In this case, the raw LF $R$
will need to be corrected upward to obtain $T$.  Since such
nondetections are also enumerated by artificial-star experiments, they
can be incorporated into the transition matrix $A$ (each column of $A$
will sum to some value less than 1.0).

\subsection{Comparison with previous work}

GYSB\markcite{GYSB} found that their WFPC2 starcounts in the M15 cusp
needed to be corrected {\it downwards}:\ they counted more stars than
were actually present, down to $V=19.$ They attributed this effect to
the fact that many of the ``stars'' they measured with magnitudes just
above the limit were actually blends of two or more fainter stars,
whose individual magnitudes were below the limit.  We see no such
effect (i.e., our counts need a slight upward correction, at the same
magnitude limit).  This difference most likely arises from the FOC's
higher resolving power:\ many of GYSB's blends should be separated on
our own images, and thus resolved as individual stars in our star
list.

To make a quick check of this hypothesis, we simulated the WFPC2's
resolution, by assuming that any artificial star with a close real
neighbor (within 0\secspt05) would have been blended with that
neighbor in GYSB's images.  We then recomputed completeness
statistics, using the {\it combined} magnitude of the artificial star
and its neighbor as the blend's recovered magnitude.  Enough
artificial stars were ``promoted'' above $V=19$ by this procedure that
the completeness at that limit climbed above $100\%$---in other words,
we recovered more stars with $V<19$ than we originally added, exactly
as seen by GYSB.

(The completeness calculated this way is not exactly the same as
GYSB's result, but we are not trying to duplicate the details of their
work here---just identifying the source of the difference between our
value of the completeness and theirs.)

\section{Star-count analysis}

\subsection{The cluster center}

Standard techniques for finding centers of clusters from star counts
rely on finding the stars' center of symmetry, in one way or another
(see\markcite{cal93} Calzetti \etal\ 1993, \markcite{me95}Sosin \&
King 1995, \markcite{GYSB}GYSB).  Large sections of the FOC images are
covered by saturated stars that are asymmetrically distributed around
any prospective center, making a measurement of the center's position
impossible.  We therefore use the position found by GYSB from their
Planetary Camera image, which appears at pixel coordinates (284, 150)
in the uppermost of the two FOC pointings that make up the field shown
in Fig.\ 1.  The position of the center is also marked in Fig.\ 3.

\subsection{Binned star counts}

Since the stellar surface density---and thus the crowding---changes
dramatically over the inner $10''$ of M15, the limiting magnitude of
our star counts varies widely with position in the central field.  At
its edge, $\sim10''$ from the cluster center, we can accurately count
stars down to nearly $V=23,$ while in the innermost arcsecond,
crowding limits the completeness of the counts to $V \simeq 20.$ (For
comparison, GYSB's limiting magnitude at the center was also $V \simeq
20,$ although their completeness was less than 50\% in the innermost
$0\secspt5$.  They did not attempt to push the counts to fainter
magnitudes at larger radii.)

For the rest of \S5, we divide the counts in the central field into
three magnitude groups.  The brightest (``turnoff'') sample contains
all stars with $18.25 \le V < 19.75$, the second (``middle'') group
has $19.75 \le V < 21.25,$ and the third (``faint'') group has $21.25
\le V < 22.75.$ We require a completeness of $>50\%$ for star counts
to be considered believable.  This implies that the turnoff group can
be counted all the way into the center; stars in the middle group are
countable in to $\sim 1''$, while the faint group can be counted only
outside of $\sim 3''$.  Within these ranges in radius, and before
correction for incompleteness, there are 577 stars in the turnoff
group, 1222 stars in the middle group, and 539 stars in the faint
group.

If we assume an apparent distance modulus of 15.40 (DH93), the
Bergbusch \& VandenBerg (1992)\markcite{BV92} isochrone discussed
above tells us that our ``middle'' group consists of main-sequence
stars with masses $0.67\Msun < M < 0.77\Msun$; for the ``faint''
sample, the mass range is $0.55\Msun < M < 0.67\Msun.$ The ``turnoff''
sample consists of stars with masses $\sim0.8\Msun$; the range in mass
of stars in this bin is much smaller than in the other two.

Figure \ref{BinnedCounts} shows the completeness-corrected star counts
in these three magnitude ranges, binned in radius.  The raw
(uncorrected) star counts are also shown, as the lines without
circular points.  To correct the raw counts for incompleteness, we
used the Drukier method to compute a transition matrix for each radial
bin and for each magnitude range, and corrected the counts
appropriately (see \S 4.3).  (We subdivided the counts into 0.5-mag
bins in the construction of $A$, $R$, and $T,$ and tabulated at least
1 mag on either side of each sample, to account for stars that jumped
into and out of the magnitude range.)  Note that the raw counts are
occasionally greater than the corrected counts, as a result of
bin-jumping.  The corrected binned counts are also given in Table 1.
(The $r$'s given in that table are the radii of the middle of each
annulus in which the stars were binned.)

As has been found by several previous investigators, the counts of
``turnoff'' stars in Fig.\ \ref{BinnedCounts} climb as a power law
with slope $\sim0.7.$ Whether they level off into a flat,
constant-surface-density core within $2''$ will be discussed further
below; for now, we note that both a pure power law and a $\sim 1''$
core appear to be equally plausible, while a $2''$ core appears much
less likely.  The counts of stars in the two fainter groups---never
before resolved---show similar power-law behavior.  Since the
maximum-likelihood method presented in the next section provides more
robust estimates of cusp slopes than would a fit to the binned counts,
we defer discussion of these results until later.

\subsection{Maximum-likelihood analysis of the density profile}

A disadvantage of binned counts is that information is lost in the
binning process.  Since we have positions and magnitudes for each
star, it is desirable to use all of this information simultaneously in
finding a ``best-fit'' profile.

The maximum-likelihood (ML) method for density-profile fitting, which
does use all of this information, was originally described by
Sarazin\markcite{sar80} (1980).  If the counts of objects are
complete, then any prospective surface-density model $f(r)$ can be
regarded as a probability distribution of the position of any
particular object, when $f$ has been normalized to 1 over the area on
the sky covered by the image.  The log-likelihood of seeing the data
(the $N$ counts), given $f(r)$, is the sum
\begin{equation}
\log L = \sum_{j=1,N} \log f(r_j),
\end{equation}
where $r_j$ is the projected radius of star $j$.  If we have a family
of density models, then that model $f$ that maximizes the likelihood
function $L$ is taken to be the ``best-fitting'' model.  In
particular, it has become common to fit the centers of globular
clusters with the family of functions
\begin{equation}
f(r;\alpha,r_c) = {1 \over {[1 + C_\alpha (r/r_c)^2]^{\alpha / 2}}},
\end{equation}
where the parameters $\alpha$ and $r_c$, respectively, are the 
asymptotic slope of the density cusp (for $r \gg r_c$), and the core
radius, at which the projected density falls to half its central value.
The constant $C_\alpha$ is chosen to make this true for any $\alpha$:
\begin{equation}
C_\alpha = 2^{2/\alpha} - 1.
\end{equation}

It is tempting to regard the likelihood function $L(\alpha,r_c)$ as a
probability distribution of the parameters, and to infer confidence
limits on these parameters directly from $L$.  Strictly speaking, $L$
cannot give us a distribution of the parameters unless we take a
Bayesian approach:\ if we assume a uniform prior distribution of
models over some region in $(\alpha,r_c)$ space, then $L$ is identical
to the probability distribution of $\alpha$ and $r_c$ over that region
(after normalization).  There could, however, be other reasonable
assumptions for a prior distribution: for example, we might regard a
core radius of 0\secspt1 to be just as likely as a core radius of
$1''$, in which case we should choose a prior that is distributed
uniformly in $\log r_c$.  For the purposes of this paper, though, we
will assume the simpler case of a uniform prior in $\alpha$ and $r_c$.

\subsection{Incorporating incompleteness into the ML method}

The presence of incompleteness implies that the star counts actually
observed will match a distribution $g(r)$ that has been adjusted for
incompleteness, rather than the intrinsic distribution $f$.  We should
therefore fit the observed counts to $g$, and choose the $f$
corresponding to the best-fitting $g$ as our estimate of the density
profile.

Why not correct the counts themselves, as we did with the binned
counts, and compare those corrected counts with the $f$'s directly?
The reason is that the statistics underlying the ML method depend upon
the individual stellar positions being independent events, whose joint
probability forms the likelihood function.  Fitting to corrected
counts would violate this assumption, since more than one corrected
count would arise from a single detection.  (For example, in a part of
the image where the completeness was 0.5, $L$ would be the probability
of finding two stars at each detected stellar position.  This is not
what actually happened; moreover, it gives a high weight to parts of
the image that were measured poorly.)

Now, consider for a moment that we have a set of $f$'s that are of
interest---e.g., the set of cluster profiles defined by Eq.\ (2).  To
generate the $g$'s actually used in the ML fit, we must multiply each
$f$ by some factor at each radius to account for incompleteness.  But,
as discussed in \S 4.2, the process of finding the appropriate factors
for some hypothetical model is not straightforward, since our
artificial-star experiments give us those correction factors for our
actual image, rather than for any hypothetical image.

Nevertheless, we can apply those same correction factors to our
hypothetical models {\it if} we can show that the incompleteness was
caused by brighter stars, rather than crowding of faint stars, so that
the completeness factors would be independent of the $f$ we are
fitting.  (The details of how the completeness was tabulated as a
function of radius will be given in the next section; here we are more
concerned with the reasons a particular star was or was not
recovered.)

To do so, we looked at the artificial-star results for turnoff stars
more closely.  We noted that in the innermost $3''$, only $\sim1/3$ of
the non-recovered artificial stars had real stars in their own
magnitude range as close neighbors.  This suggested that only
$\sim1/3$ of the incompleteness of turnoff stars was due to blends
with other turnoff stars, and that the remaining $2/3$ was due to
somewhat brighter subgiants.  A visual inspection of the locations of
the non-recovered artificial stars led us to a similar conclusion.
(Note that even more obvious areas of saturation had already been
excluded before this stage.)

For artificial stars in the other two samples (the ``middle'' and
``faint'' groups), blending with stars in their own magnitude range
was quite rare.  The incompleteness for these fainter main-sequence
stars was caused entirely by turnoff stars and subgiants.

Since most of the incompleteness was due to brighter objects, we
proceeded under the approximation that the incompleteness fraction
measured at a given radius in our artificial-star experiments was
appropriate for all prospective profiles (Case I from \S 4.2).

\subsection{Radial dependence of the completeness fraction}

We used the artificial-star results (\S 4.1) to find a completeness
ratio $\psubm(r)$ for each magnitude bin $m$.  We defined this as
$$ \psubm(r) = {{\hbox{\rm number of stars that are detected, in the
                       magnitude range, and at radius } r}\over 
           {\hbox{\rm number of stars actually present, in the
                       magnitude range, and at radius } r} }.$$
To fit a density profile, then, we used the ML method (\S 5.3) to find
the best-fitting $g(r) = \psubm(r) f(r),$ where $f$ is chosen from the
set of functions $f(r;\alpha,r_c)$ defined in Eq.\ 2.  As discussed in
\S 4.2 and justified in \S 5.4, $\psubm$ was a function only of radius, and
not of the density itself.

We used the same magnitude ranges in the ML fit as were defined in \S
5.2, but we needed to tabulate each $\psubm$ as a function of $r$
without an arbitrary binning in radius.  From our completeness
experiments, we had a database of $\sim6 \times 10^4$ artificial
stars.  At the radius $r$ of each real star, we found the $N\subr{a}$
artificial stars nearest in radius to $r$ (we chose $N\subr{a}$ to be
1000 for the turnoff group, and 3000 for the fainter groups).  The
positions of the $N\subr{a}$ artificial stars in this subsample
defined an annulus, with midpoint $r$ and width $\sim 1''$.  Next, we
constructed the transition matrix $A$ for the subsample (see \S 4.3).
We then tabulated the local raw LF vector $R$ for the real stars that
fell in the same annulus as the subsample, and applied $A^{-1}$ to $R$
to obtain the local true LF $T$.  Finally, we found each $\psubm(r)$
as the ratio of the sums of the appropriate elements of $R$ and $T$.
This method ensured that for each real star the $\psubm$ term used in
the likelihood function was the most appropriate value at its
position; no radial binning was necessary.  Note also that $\psubm$
depends on the number of {\it real} stars, not just the artificial
stars, since the artificial-star results must be weighted by the local
LF in the computation of the overall completeness.

For stars at small radii (in the turnoff sample), for which there are
fewer than $N\subr{a}$ artificial stars within $r$, we replaced the
annulus with a circle of radius $2r$.  This allowed us to track the
changes in $\psubm$ within the innermost arcsecond.  The circle was never
allowed to have fewer than 250 artificial stars in it.  We also
ignored the region within 0\secspt3 of the cluster center, because the
precise position of the center is not known, and because small-number
statistics begin to dominate $\psubm.$

The computed $\psubm(r)$ is quite noisy.  One reason is that the
number of real stars within the annulus varies with $r$ (since the
local raw LF $R$ is used in the computation of $\psubm$, noise in $R$
leads to fluctuations in $\psubm$).  A less important source of noise
is the fact that the off-diagonal elements in the transition matrix
$A$, which correspond to rare, large jumps in magnitude, are not well
sampled.  The high-frequency fluctuations in $\psubm$ lead to problems
in the ML fit, since the product $\psubm(r) f(r)$ must be normalized
over the field of view; the noise makes the integrator take far too
many steps.  We therefore applied a second-order Savitzky--Golay
smoothing filter (Press\markcite{nr} \etal\ 1992) with a smoothing
length of 100 stars to the raw $\psubm(r)$, to generate the $\psubm$
actually used in the profile fitting.  (The second-order
Savitzky--Golay filter preserves second-order moments, so that peaks
or troughs in $\psubm$ will not be smoothed out.)

The smoothed $\psubm(r)$ for each of the three stellar samples is
shown in Figure \ref{RadialCompleteness}.  (In this section we
concentrate on the turnoff sample; we will discuss the other two in \S
6.)

Note that $\psubm(r)$ still fluctuates on radial scales shorter than
the variations in the density profile.  More smoothing would remove
the remaining fluctuations, but not without losing resolution of
features in $\psubm(r)$ near the cluster center; we therefore decided
not to smooth $\psubm$ any further.  We did check the behavior of the
ML fit with greater smoothing, and found that the results were
essentially identical to what we will present, except that the
smoothing made the measured core radius artificially large.

(An alternative method, suggested by the referee, would be to fit an
analytic function to $\psubm$.  The danger in such a method is that
the assumed functional form itself might determine the behavior of
$\psubm$ at small radii, and thereby influence whether we deduce the
presence of a small core.  For example, the analytic function used
by GYSB [their Eq.\ 2] {\it must} approach $1 + C_1 + C_2$ near $r=0$.
The values of $C_1$ and $C_2$ are set by fits at $r > 1\secspt5$,
where the crowding conditions are quite different.)

The completeness of the turnoff sample is $\sim1$ over most of the
central field.  It drops to $\sim0.5$ only in the central arcsecond.

\subsection{The density profile of the turnoff sample}

We then fit the star counts in the turnoff sample to the family of
$f$'s defined in \S 5.3, and obtain the log-likelihood function shown
in Figure \ref{TurnoffLLF}.  The contour lines in Fig.\
\ref{TurnoffLLF} are plotted in steps of 1, so that a model that lies
on the $-2$ contour is one-tenth as likely to produce the observed
counts as a model that lies on the $-1$ line.

The likelihood function reaches its maximum at a cusp slope of
$-0.70$, and at a very small core radius.  However, models with core
radii up to $\sim1''$ yield fits of essentially the same quality as
the maximum-likelihood model; taken at face value, the 95\% upper
limit on $r_c$ is $\sim1$\secspt3.  Assuming a flat prior, the
one-sigma uncertainty of the slope $\alpha$ is $\sim0.05$, and its
95\% confidence limits are $\pm 0.09$.  This result is fairly
consistent with previous values: GYSB's result was $-0.82$ for a
similar sample, with a 95\% confidence limit of $\pm 0.12$.

It is important to realize that the cusp's rise in the inner $2''$ is
almost entirely due to the sharp drop in the completeness fraction
$p$, and not to the uncorrected stellar density, which is roughly
constant within $r=2''$.  (Again, GYSB's study is similar:\ for $V<20$
their uncorrected density is flat within $r=3''$.)  A better
determination of an upper limit on $r_c$ would require extensive and
difficult simulations (see \S 4.2).  Nevertheless, we believe that our
upper limit on $r_c$ is unlikely to be significantly in error,
although the effect of properly accounting for faint-star blends would
be to make it somewhat larger (perhaps $\sim1\secspt5$).  In any case,
our results are not consistent with the presence of a $2\secspt2$
core, as claimed by Lauer\markcite{lau91} \etal\ (1991).  We do
caution that the interpretation of maximum-likelihood fits to
completeness-corrected star counts is not as straightforward as one
might think.

\subsection{Fainter stars in the cusp}

The density profile of lower-mass stars in the central cusp of a
globular cluster has never been observed.  While the cusp is usually
assumed to be the result of gravothermal core collapse, the presence
of a black hole with a mass of $\sim10^3\Msun$ has never been ruled
out.  Observations of lower-mass stars can test whether the cluster
potential has the radial dependence predicted by core-collapse or by
black-hole models.

In comparing cusp slopes of different stellar groups, we must choose a
consistent radial range over which to measure the slopes.  We might
choose a range over which all groups are at least 50\% complete (as in
Fig.\ 6).  This is a common requirement; it is intended to prevent
small systematic errors in artificial-star experiments from dominating
the corrected counts.  However, in our case, we are interested in the
{\it variation} of the corrected density with radius.  This variation
depends on $\dpmdr$ as well as on $\psubm$ itself, implying that
systematic errors in the former can affect the deduced cusp slope just
as much as systematic errors in the latter.  For example, if there
were multiple causes of incompleteness whose relative importance
varies with density (and, therefore, position), and if we did not
account for one of those effects correctly, then our measurement of
the cusp slope would be strongly influenced by the completeness
corrections, despite the fact that $\psubm$ was always over 50\%.

This observation suggests that we set an upper limit on $| \dpmdr |$, in
addition to our $\psubm$ requirement, when measuring cusp slopes.
But, since $\dpmdr$ is difficult to measure directly, we take a less
formal approach, and simply choose to measure slopes in regions where
$\psubm$ varies reasonably slowly with $r$, for each group (see Fig.\
7, and ignore the small-scale fluctuations).  Typically, $c$ decreases
slowly with decreasing $r$ at large radii, and then drops dramatically
at some radius---near $1''$ for the turnoff group, and $2''$ for the
middle group.  For the faint group, $\psubm$ varies rapidly with $r$
over the entire central field, meaning that the accuracy of its radial
profile depends strongly on the absence of any systematic errors in
the artificial-star procedure.  (The {\it total} number of faint stars
in the outer part of the central field is better known, since it
depends only on an average $\psubm$.)

The upper panel of Figure \ref{FaintLLF} shows the likelihood function
of cusp slopes for the three stellar samples.  To allow us to compare
the slopes of the turnoff and middle groups, and to minimize the
influence of any flat core on these fits, we ignored stars at radii
less than $2''$, and fit the counts with essentially pure power laws
($r_c \rightarrow 0$ in Eq.\ 2).  (For the faint group, the inner
radial limit was $r=3''$, the radius where its completeness climbs
above 50\%.)  The lower panel of Fig.\ \ref{FaintLLF} shows the
likelihood functions in a cumulative form.  If we assume a uniform
prior (see \S5.3), this panel shows the probability of the cusp slope
being flatter than a given value.

The solid line refers to the turnoff sample; its best-fit cusp slope
here is $-0.64 \pm 0.08$.  The dotted line is the middle group; its
slope is $-0.56 \pm 0.05$.  (The value for the turnoff group is
flatter than the value found in the previous section, because we are
using only stars with $r>2'',$ rather than $r>0\secspt3$.  In Fig.\ 6,
one can see that the profiles of both groups steepen a bit near
$r=2''$.  If we set the inner radial limit at $1''$ instead, these
slopes become $-0.67$ and $-0.64$, respectively.)  The dashed line in
Fig.\ 9 is the faint group; at face value it appears to decline very
steeply ($-0.8 \pm 0.1$).  However, this last slope is dominated by
the radial variation of the completeness correction, and probably
should not be trusted.

What do models of the density cusp predict for the variation of cusp
slope with stellar mass?  Let $-\beta$ be the cusp slope in {\it
space} density, so that the (negative) slope of the surface-density
profile is $\alpha = \beta - 1$ (for $\beta > 1$).  Models of clusters
in the process of core-collapse (Cohn\markcite{hc85} 1985) predict a
strong variation of $\beta$ with mass:
$$ \beta\subr{cc} = 1.89 \biggl({m \over m\subr{max}}\biggr) + 0.35, $$
where $m\subr{max}$ is the mass of the stars that dominate the cusp.
These models also predict a space-density slope of $\beta \simeq 2.23$
for the dominant stars, which corresponds to a surface-density slope
of $\alpha \simeq 1.23$ (Cohn\markcite{hc80} 1980).  The potential well
in the cusp varies as $\Phi(r) = GM/r \propto r^{2-\beta} \propto
r^{-0.23},$ assuming that it is dominated by the heaviest stars.

For the $0.8\Msun$ turnoff stars to have $\alpha \simeq 0.64$ in a
collapsing model, we must hypothesize the existence of a population of
dark objects with masses near $1.2\Msun$, or some combination of
populations whose effect is similar.  Such a model would predict
slopes $\alpha \simeq 0.48$ and $0.32$ for our two fainter groups,
taking their mean masses to be $0.7$ and $0.6\Msun$.  An isothermal
cusp in thermodynamic equilibrium, which might be a more appropriate
model of the post-core-collapse phase, would predict slightly flatter
slopes ($0.44$ and $0.23$, respectively).

Black-hole models, on the other hand, predict far less variation of
cusp slope with stellar mass (Bahcall \& Wolf\markcite{bw77} 1977),
since the potential well is much steeper in the BH case:\ $r^{-1}$
rather than $r^{-0.23}$.  The BH models predict $\alpha = 0.75$ for
the most massive stars in the cusp (the turnoff stars, in this case),
with a slow variation of slope with stellar mass, down to $\sim0.5$
for very-low-mass stars.  Over the range of masses covered by our
observations, the difference in slope should be negligible; all three
groups should have slopes near 0.7.

The turnoff group's slope is consistent with either the core collapse
model (assuming heavy remnants are present), or the black hole model
(the slope is within $1.0$ or $1.4\sigma$ of $-0.75$, depending on
which radial range is adopted).  The middle group's slope measured
here, though, falls between the predictions of the two models.  From
Fig.\ 9, the probability of the slope being $\le 0.48$ (the
core-collapse prediction) is about $7\%$; while the probability of its
being $\ge 0.7$ is only about $1\%$.

Overall, the core-collapse model is somewhat more consistent with
these data than is the black-hole model, although neither can be
considered a very good fit.  However, it should be noted that the cusp
slopes seen by Chernoff \& Weinberg (1990)\markcite{cw90} in their
Fokker--Planck simulations did not always agree with the
\markcite{hc85}Cohn (1985) prediction, so true evolving core-collapse
models may match these observations more closely.

\section{Mass segregation}

From the $V$ magnitudes of stars in the central and $r=20''$ fields,
we derive the luminosity functions shown in Figure \ref{LumFunctions}.
(The uppermost LF refers to the same stars that were discussed in \S
5, but we now plot them as a luminosity function, instead of a
surface-density profile.)  The LFs have been corrected for
incompleteness and ``bin-jumping'' by the Drukier matrix method (see
\S4.3).  We plot the central field LF down to $V = 22$, for $3'' <
r < 8''$.  Note that the {\it total} number of faint stars in the
indicated radial range is not affected by the uncertainties in $\dpmdr$
that made the determination of their radial profile difficult, so the
caveats from \S5.7 do not apply to Fig.\ 10 (or to Fig.\ 11 below).

We next convert magnitudes to masses using the [Fe/H] $= -2.26,$ $t =
15$ Gyr isochrone of\markcite{BV92} BV92 (the isochrone plotted in
Fig.\ \ref{OuterCMD}).  The best choice of mass-luminosity relation
is a matter of some debate, but we are concerned with stars with
masses above $0.5 \Msun$, for which the differences between M--L
relations are minor.  We choose the BV92 isochrone because it
reproduces the magnitudes and colors near the main-sequence turnoff
quite well\markcite{DH93} (DH93); more recent calculations that fit
the lower main sequence have not attempted to reproduce the turnoff.

We plot the resulting mass functions in Figure \ref{MassFunctions},
along with the MF recently measured by Piotto \etal\ (1996)
\markcite{gp96}in a field $5'$ from the center of M15, using the
WFPC2 on \hst.  The two FOC MFs refer to the left-hand scale, while
the WFPC2 MF refers to the right-hand scale.  The stellar density is
much lower in the $r=5'$ field; the vertical position of the WFPC2 MF
has been chosen to match the $r=20''$ MF at the bright end, so that
the two may be easily compared.  The MFs are also given in Table 2.

The $r=20''$ and $r=5'$ (WFPC2) MFs clearly show substantial mass
segregation; the former is much flatter than the latter.  (Note that
differences between the $V$ filters used in the WFPC2 and FOC are
minor, and certainly could not account for the difference in MFs.)
Their difference is in the direction one would expect from two-body
relaxation.  The innermost MF continues this trend, with even fewer
low-mass stars.

A quantitative, model-independent measurement of the degree of mass
segregation is difficult: the three MFs do not cover the same mass
range, and none of the MFs are well fit by power laws.  Nevertheless,
we can compare rough power-law slopes over various ranges of mass.
Over the range from the turnoff ($0.78\Msun$) to $0.45
\Msun$, the $r=5'$ WFPC2 MF is best fit by a power-law with a slope $x
= 1.00 \pm 0.25$ (where the Salpeter slope is 1.35); this power-law
slope is in excellent agreement with the results of DH93 over a
similar mass range.  The $r=20''$ MF is best fit by $x = -0.75 \pm
0.26$ over the same range; the two are therefore different at the
$\sim5\sigma$ level.

Considering only the range in mass from $0.78$ to $0.60 \Msun$, for
which data in all three radial bins are available, we measure a slope
$x = 2.1 \pm 1.0$ for the $r=5'$ field, $x = 0.2 \pm 0.7$ for the
$r=20''$ field, and $x = -2.2 \pm 0.4$ for the $3'' < r < 8''$ sample.
In this case, the last two slopes differ at the $3\sigma$ level.

Computational simulations of globular-cluster evolution have usually
assumed a power-law mass function, so a comparison with theory also
requires us to choose a mass range.  Grabhorn\markcite{gra92} \etal\
(1992) constructed a Fokker--Planck model of M15 with a global MF
slope of $x\subr{global} = 0.9$---quite close to the measured value at
$r = 5'$ over the first of the two mass ranges considered above.
Figure 8 of Grabhorn \etal\ shows the radial variation of the MF slope
at the time of deep core collapse in their model; from that plot, $x =
1.2$ at $r = 5'$, $x = 0.1$ at $r = 20''$ and $x = -0.4$ at $r \simeq
5''$.  The differences between these slopes appear to be smaller in
the model than in our observations.  If this particular ``snapshot''
of the model is appropriate for comparison with M15 at the present
epoch, then the model is underestimating the degree of mass
segregation in the cluster.

(Note that here we are comparing the {\it overall} MF of each field
with the others.  Within the central cusp itself, the Fokker--Planck
models {\it overestimate} the radial variation of the MF [\S5.7].)

Drukier\markcite{dru95} (1995) has recently constructed an evolving
Fokker--Planck model of the cluster NGC 6397, which he could compare
with the observed surface-density profile and mass functions at many
stages of the calculation.  The data presented in this paper should be
useful for similar comparisons with evolutionary models of M15.

\section{A King--Michie model of M15}

Lacking the resources to run a Fokker--Planck or $N$-body model of
M15, we instead constructed a multimass King--Michie model.  Such
models do not include many of the physical processes that make M15
such an interesting object to study (e.g., core collapse), but are the
simplest models that can give a somewhat realistic prediction of the
variation of the mass function throughout the entire cluster,
including the region near the tidal boundary.  They are also easy to
calculate, and are a first step with which more sophisticated models
can be compared.

The hallmark of the King--Michie model is its ``lowered Maxwellian''
distribution function (DF); the DF goes to zero at the cluster's
escape energy.  King\markcite{irk65} (1965) showed that this DF
approximates the steady-state solution of the Fokker--Planck equation;
the first actual models were calculated by Michie \&
Bodenheimer\markcite{mic63} (1963) and King\markcite{irk66} (1966).
The most general form of the models, which includes anisotropy of the
velocity dispersion and multiple stellar mass groups, was published
by\markcite{gg79} Gunn \& Griffin (1979).  The code used to calculate
the model presented here uses the notation and technique of Gunn \&
Griffin.

To calculate a model, we first specify a set of mass groups.  We
choose a mean mass for each group, and some constraint on the number
of stars in the group---either the total number of stars, or the
projected stellar density at some radius (which is usually chosen to
agree with an observed mass function).  We then supply the desired
core, tidal, and anisotropy radii, which are chosen so that the
model's density profile agrees with observation.  The code then
iterates for a model with the chosen set of parameters.  After the
model has been computed, we can check its predicted stellar velocities
and mass segregation against observation.

For the model presented here, we chose a tidal radius $r_t$ of 60
parsecs and a core radius $r_c$ of $0.01\> {\rm pc} =0\secspt2$; the
small $r_c$ allows the model's density cusp to continue in to small
radii.  Values of $r_c$ up to 0.1 pc lead to models with virtually the
same predictions; the only difference is that the density profile
flattens near the innermost density points.  We also used an
anisotropy radius, $r_a,$ of $22\> {\rm pc} = 7\minspt3$ (the stellar
velocities become increasingly radial beyond $r_a$).  We were unable
to fit the density profile well with any model with isotropic
velocities; the surface-density profile of the best-fitting isotropic
models fell below the observed profile in the ``shoulder'' around
$r=20''$.

The model's mass function consisted of 15 groups.  The first 13 groups
were constrained to agree with the observed WFPC2 mass function at $r
= 5'.$ Group 14 contained white dwarfs, with a mean mass of 0.55
$\Msun$.  The number of WDs was chosen so that $\sim 20\%$ of the
cluster mass was in WDs, in rough agreement with various other models
that have been calculated (e.g., Meylan\markcite{mm91} \& Mayor 1991).
The WD mass fraction could be varied between 10\% and 30\% without
affecting the model dramatically.

Group 15 contained massive, dark stellar remnants---neutron stars or
heavy WDs.  Since these stars are the heaviest objects in the model,
they concentrate strongly toward the cluster center, and control the
inner density profile of the luminous giants.  As in the core-collapse
models discussed in the previous section, the {\it individual} mass of
the heavy remnants determines the slope of the density cusp of the
turnoff stars, while their {\it total} mass determines the cusp's
radial extent.  To match the cusp, then, we put $\sim1800$ $1.5 \Msun$
stars into Group 15, which constituted $0.78\%$ of the total cluster
mass of $3.4 \times 10^5 \Msun$.  (The cusp extends to about $10''$,
although the heavy remnants dominated the total mass density out to a
radius of only $2''$ or so.)  It is interesting that the best-fit
remnant mass is quite close to the Chandrasekhar mass ($1.44 \Msun$).

(Some other pre-\hst\ cluster modeling has taken the approach of
varying the MF power-law slope until a model can be found that agrees
with observations; since the number of remnants in those models is set
by an extrapolation of the MF slope, the procedure is essentially
similar to ours.  In our case, since we now know the main-sequence MF
from observation, we must adjust the number of remnants
independently.)

One might criticize the model on the grounds that it is rather {\it ad
hoc}.  For instance, the numbers of WDs and heavy remnants are not
necessarily consistent with an extrapolation of the chosen MF.
Furthermore, the anisotropy radius was regarded as a free parameter;
another procedure might have been to {\it require} $r\subr{a}$ to be
equal to the radius at which the relaxation time is equal to the
cluster's age (the two did turn out to be within about a factor of
two of each other).  However, our goal here is to see whether any
King--Michie model can adequately {\it describe} the observed state of
the cluster; we leave the interpretation of any successful fits for
another time.

The comparison of the model with observed quantities is shown in
Figure \ref{KingModel}.  (The model parameters and mass functions are
listed in Tables 3a and 3b.)  The lower right panel shows the mass
function at $r = 5'$ as measured by Piotto\markcite{gp96} \etal\
(1996) and as it appears in the model output.  Their similarity
confirms that the iteration procedure successfully matched its
constraints.  (Note that the model's mass function does not {\it
exactly} match the measured MF.  The MF used in the model was rounded
off to even quantities, to avoid unnecessary scatter in the plotted
points.)

The upper left panel of Fig.\ \ref{KingModel} shows the projected
surface density profile.  The line is the model output, which was
matched to the plotted points.  The heavy points are the turnoff
stars' profile from Fig.\ \ref{BinnedCounts}, and the lighter points
are ground-based observations from Lugger\markcite{lug87} \etal\
(1987) and King\markcite{irk68} \etal\ (1968).  The lower part of this
panel shows the logarithmic difference between the surface-density
data and the model.

The upper right panel of Fig.\ \ref{KingModel} compares the predicted
and observed inner mass functions; the plotted points are identical to
those shown in Fig.\ \ref{MassFunctions}.  The solid lines are the
model predictions.  The predicted MF for $3''<r<8''$ has been averaged
over the area of the central field, so that it can be compared
directly with the observed MF.  Unlike the Fokker--Planck models, the
King--Michie model predicts {\it stronger} mass segregation than is
actually observed:\ both model lines go down with decreasing mass
faster than the data.  (The amount of disagreement is somewhat
sensitive to details of the assumed $r=5'$ MF, in particular, to the
slope of the MF's rise near $\log m = -0.2$.  But no such adjustments
could make all of the predicted MFs consistent with observation.)

Finally, the lower left panel of Fig.\ \ref{KingModel} shows the
model-predicted velocity dispersion profile, along with the binned
velocity dispersion profiles of Peterson, Seitzer, \&
Cudworth\markcite{psc89} (1989, plotted as squares) and
Gebhardt\markcite{geb94} \etal\ (1994, plotted as triangles).  (The
questionable innermost point of the Peterson \etal\ profile is not
shown.)  Like other high-concentration King--Michie models, the model
shown here is nearly isothermal in its center, and does not reproduce
the rise in M15's velocity dispersion in the inner $30''$.

The Grabhorn\markcite{gra92} \etal\ (1992) Fokker--Planck model of M15
{\it does} reproduce the velocity profile, at least more closely:
core-collapse models have a power-law dependence of the velocity
dispersion on radius, with logarithmic slope $-0.23$.  The Grabhorn
model also fit the cluster with a much larger fraction of heavy
remnants (9\% of the cluster mass, rather than 0.8\%), which brought
their {\it central} velocity into agreement with the accepted
observational value of $\sim14 \km \s^{-1}$ (Dubath, Meylan, \&
Mayor\markcite{dmm94} 1994).  Putting such a high fraction of heavy
remnants into the King--Michie model would allow the central velocity
dispersion to agree with observation, but not the run of velocity
dispersion with radius.  The agreement of the density profile with
observation would also be destroyed.

In summary, we can fit a King--Michie model to the observed mass
function and surface-density profile of M15, but the model does not
fully reproduce the observed degree of mass segregation.  Since King
models were already known not to fit the stellar velocities
(Grabhorn\markcite{gra92} \etal\ 1992, Gebhardt\markcite{geb95} \&
Fischer 1995), this lack of agreement is not a surprise.  However, our
model is the first to use the actual observed cluster MFs, and thus
confirms that M15 is not in dynamical equilibrium.

\section{Conclusions}

We have presented radial distributions and mass functions of
main-sequence stars in two fields near the center of M15.  The
distribution of turnoff stars in the innermost $10''$ is similar to
what has been found previously: we find a power-law density cusp, with
logarithmic slope $-0.70 \pm 0.05$ (for stars with $r>0\secspt3$).
The density continues to rise in to small radii; any
constant-surface-density core must be smaller than $1\secspt5.$

We have shown, for the first time, that fainter, lower-mass stars also
have a power-law distribution near the cluster center; we find a
logarithmic slope of $-0.56 \pm 0.03$ for a group of stars with masses
near $0.7\Msun$, over the radial range from $2''$ to $10''$.  (Over
the same range, turnoff-mass stars have a slope of $-0.64 \pm 0.08.$)
These slopes are not well matched by the predictions of the simplest
core-collapse and black-hole models.

We have also compared the mass functions in our fields with the MF in
a field $5'$ from the cluster center, and found a strong degree of
mass segregation.  Finally, we fit a King--Michie model to the
cluster's surface-density profile and outer mass function, and showed
that it predicts somewhat less segregation than we observe.  The model
also does not match the cluster's velocity-dispersion profile.

While our observations alone cannot settle the debate over the
dynamical state of M15, we hope our results will provide a valuable
set of inputs to a new generation of sophisticated and realistic
models of globular-cluster evolution.

\acknowledgements

It is a pleasure to thank Jay Anderson, Aaron Barth, Adrienne Cool,
and George Djorgovski for a number of suggestions in the course of
this project, and Charles Bartels for computer assistance.  We also
thank an anonymous referee for many helpful comments.  This work was
supported by NASA grant NAG5-1607.

\newpage

%
%
 
\figcaption[Sosin.fig1.ps]{
F480LP (FOC ``$V$'') image of a $13.6 \times 7.2$-arcsecond
region near the center of M15.  The cluster center is near the middle
of the figure.
\label{M15Inner}}

\figcaption[Sosin.fig2.ps]{
F480LP (FOC ``$V$'') image of a $7 \times 7$-arcsecond region,
20 arcsec from the center of M15.  The scale is twice as large as in
the previous figure.  The circle indicates the position of the blue
star described in \S3.2.
\label{M15Outer}}

\figcaption[Sosin.fig3.ps]{
F480LP (FOC ``$V$'') image of the innermost 3\secspt5 $\times$
3\secspt5 of M15.  The cross marks the position of the cluster center
as found by GYSB (see \S5.1).  Note the diffraction rings around the
images of stars, and the severe crowding in the innermost $\sim1''$.
Stars of various $V$ magnitudes are indicated.  The large
crescent-shaped regions are badly saturated bright stars.
\label{CenterCloseup}}

\figcaption[Sosin.fig4.ps]{
An illustration of the image reductions.  The five
panels show the same small section of the central field at different
stages of the reduction process.  From left to right: (a) in the
original image, (b) after the first pass, (c) after diffraction rings
have been removed using the first-pass results, (d) after the second
pass, (e) the final subtracted image.
\label{ReductionScheme}}

\figcaption[Sosin.fig5.ps]{ A color--magnitude diagram, in Johnson $B$
and $V$, of the $r=20''$ field.  The solid line is the 15 Gyr,
$Z=10^{-4}$ isochrone of Bergbusch \& VandenBerg (1992), and the
dashed line is the fiducial main sequence of Durrell \& Harris (1993).
The deviation of the points from the lines at the bright end is a
result of detector nonlinearity in the $B$ image (see the text).
\label{OuterCMD}}

\figcaption[Sosin.fig6.ps]{
The projected stellar density in the inner $10''$, plotted in radial
bins, in three magnitude ranges.  The circular points connected by
lines refer to the counts corrected for incompleteness.  The lines
without points show the ``raw'' counts, prior to correction.
\label{BinnedCounts}}

\figcaption[Sosin.fig7.ps]{
The completeness ratio $p_{\rm m}(r)$ for the three stellar samples.
See the text for an explanation of the fluctuations.
\label{RadialCompleteness}}

\figcaption[Sosin.fig8.ps]{
The log-likelihood of the ``turnoff'' sample ($18.25 < V < 19.75$), as
a function of the cusp slope $\alpha$ and core radius $r_c$ in
arcseconds, using the fitting function defined by Eq.\ 2.  Lighter
areas indicate a higher probability of seeing the observed counts.
Contour lines are plotted one unit apart, so that each line represents
a factor of 10 in the likelihood function.  Thus, a model with an
$(\alpha,r_c)$ pair that lies on the $-2$ contour line is one-tenth as
likely to produce the observed counts as a model whose $(\alpha,r_c)$
lies on the $-1$ line.  The point of maximum likelihood lies at
$\alpha = 0.7$ and very small $r_c$.
\label{TurnoffLLF}}

\figcaption[Sosin.fig9.ps]{
(Upper panel) The likelihood of seeing the observed counts as a
function of the cusp slope $\alpha$, for each of the three stellar
samples, using the fitting function defined by Eq.\ 2.  The solid
curve refers to the turnoff magnitude bin ($18.25 < V < 19.75$), the
dotted curve to the middle bin ($19.75 < V < 21.25$), and the dashed
curve to the faint bin ($21.25 < V < 22.75$).  (Lower panel) The same
likelihood functions converted to a cumulative probability, assuming a
uniform prior.
\label{FaintLLF}}

\figcaption[Sosin.fig10.ps]{
Luminosity functions in the central field and the $r=20''$ field.
\label{LumFunctions}}

\figcaption[Sosin.fig11.ps]{
Mass functions in the central field and the $r=20''$ field.
\label{MassFunctions}}

\figcaption[Sosin.fig12.ps]{
The King--Michie model, fit to our observations and to others from the
literature.  In all panels, points are observations, and solid lines
are model constraints or predictions.  (UL) The surface-density
profile (model constrained to fit the observations).  Inner points are
from this work; outer points are from Lugger \etal\ (1987) and King
\etal\ (1968).  (LL) The velocity-dispersion profile (model does not
reproduce the central rise in velocity dispersion).  Points taken from
Peterson, Seitzer, \& Cudworth (1989, plotted as squares), and
Gebhardt \etal\ (1994, plotted as triangles).  (LR) The mass function
at $r=5'$ (model constrained to fit the data), from Piotto \etal\
(1996). (UR) Mass functions in the central and $r=20''$ fields (model
predicts stronger mass segregation than observed, when compared with
$r=5'$ field), from this work.
\label{KingModel}}

%
%

\newpage

\begin{figure}

\vskip0.4truein
\centerline{Figure \ref{M15Inner}}

\end{figure}

\clearpage

\begin{figure}

\vskip0.4truein
Figure \ref{M15Outer}
\end{figure}

\clearpage

\begin{figure}

\vskip0.4truein
Figure \ref{CenterCloseup}
\end{figure}

\clearpage

\begin{figure}

\vskip0.1truein
\hbox{\hbox to 1.7truecm{ }
\hbox to 3.00truecm{a}
\hbox to 3.00truecm{b}
\hbox to 3.00truecm{c}
\hbox to 3.00truecm{d}
\hbox to 3.00truecm{e}}

\vskip1.0truein
Figure \ref{ReductionScheme}
\end{figure}

\clearpage

\begin{figure}
\plotone{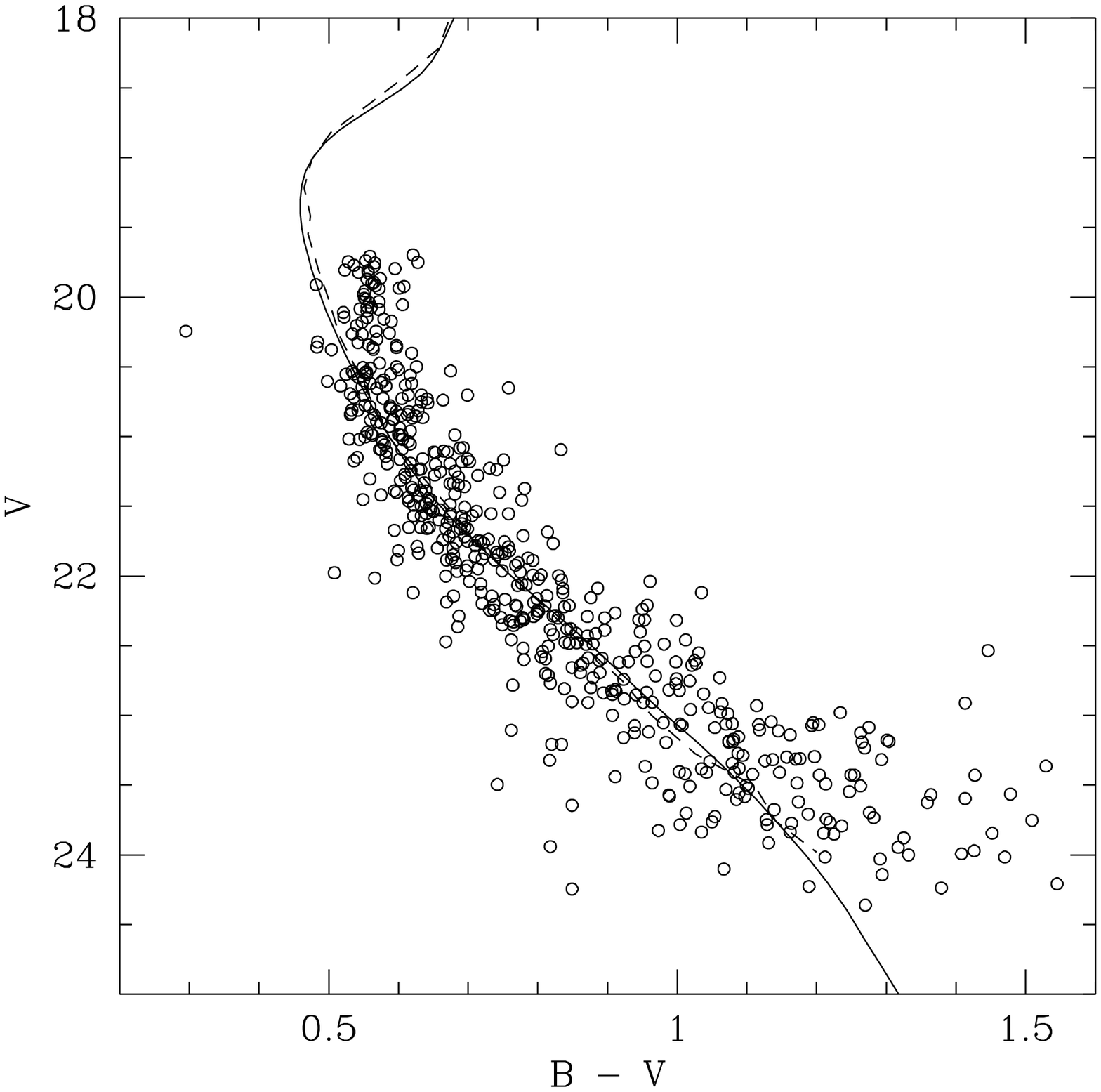}

Figure \ref{OuterCMD}
\end{figure}

\clearpage

\begin{figure}
\plotone{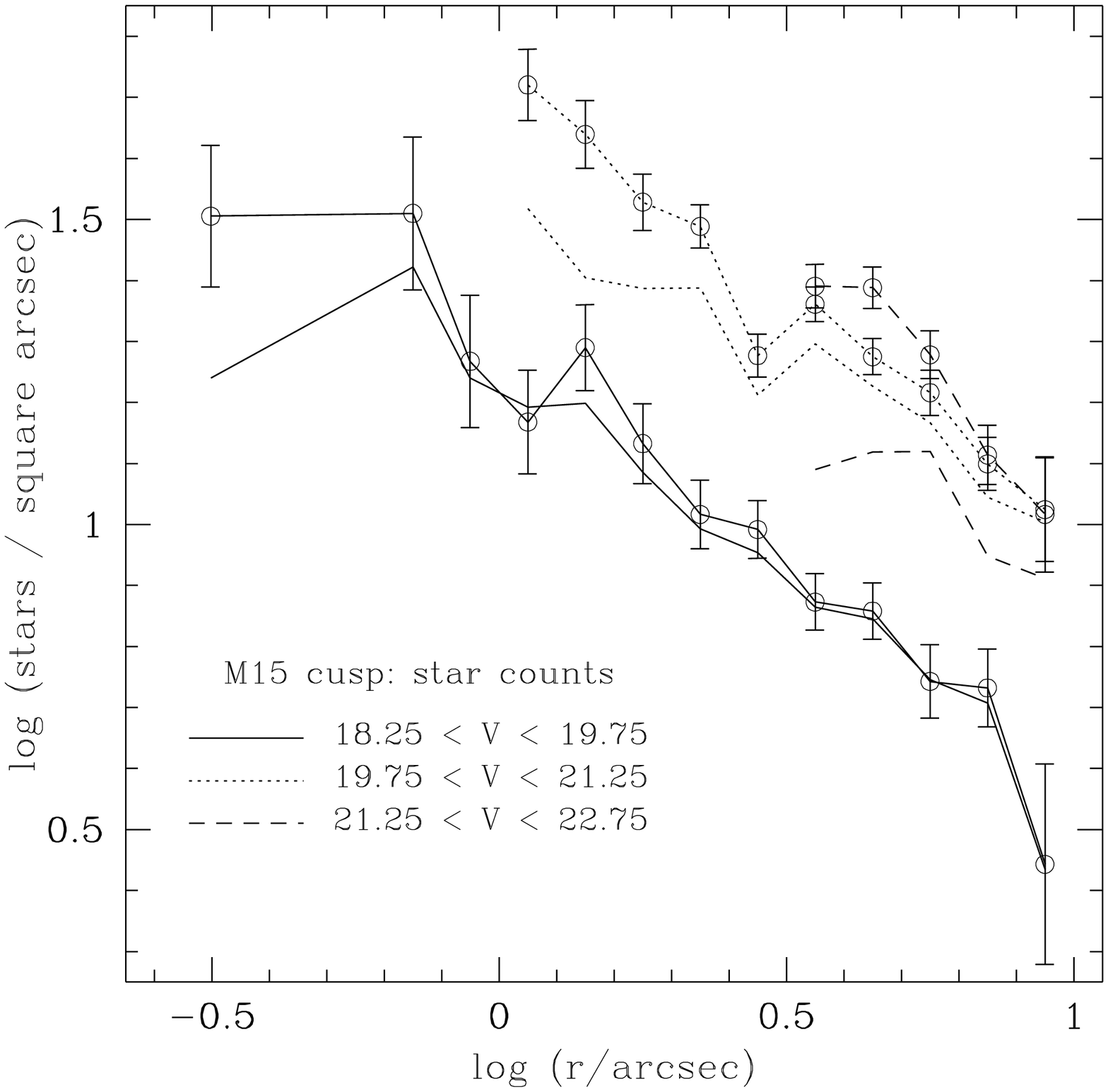}

Figure \ref{BinnedCounts}
\end{figure}

\clearpage

\begin{figure}
\plotone{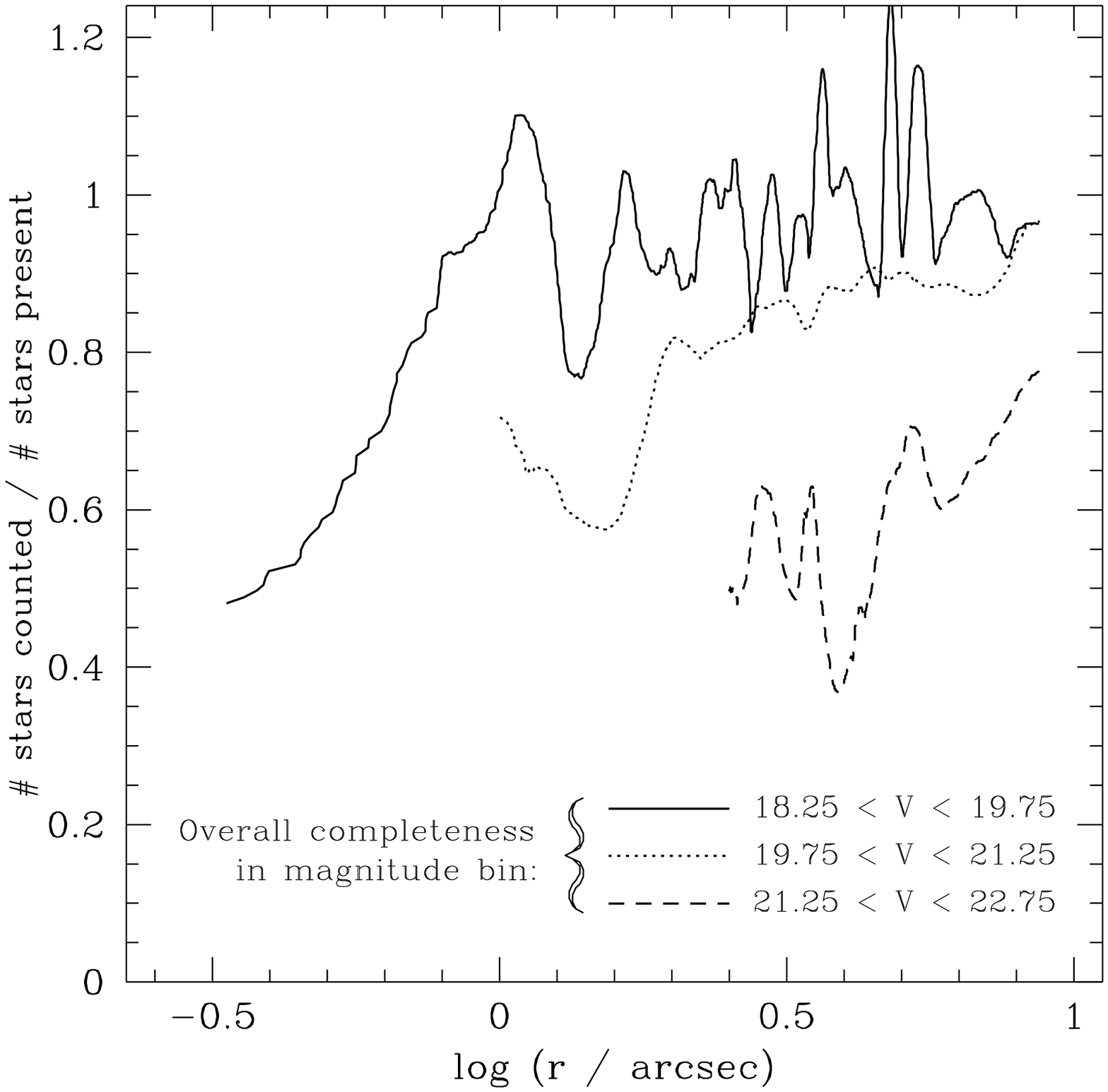}

Figure \ref{RadialCompleteness}
\end{figure}

\clearpage

\begin{figure}
\plotone{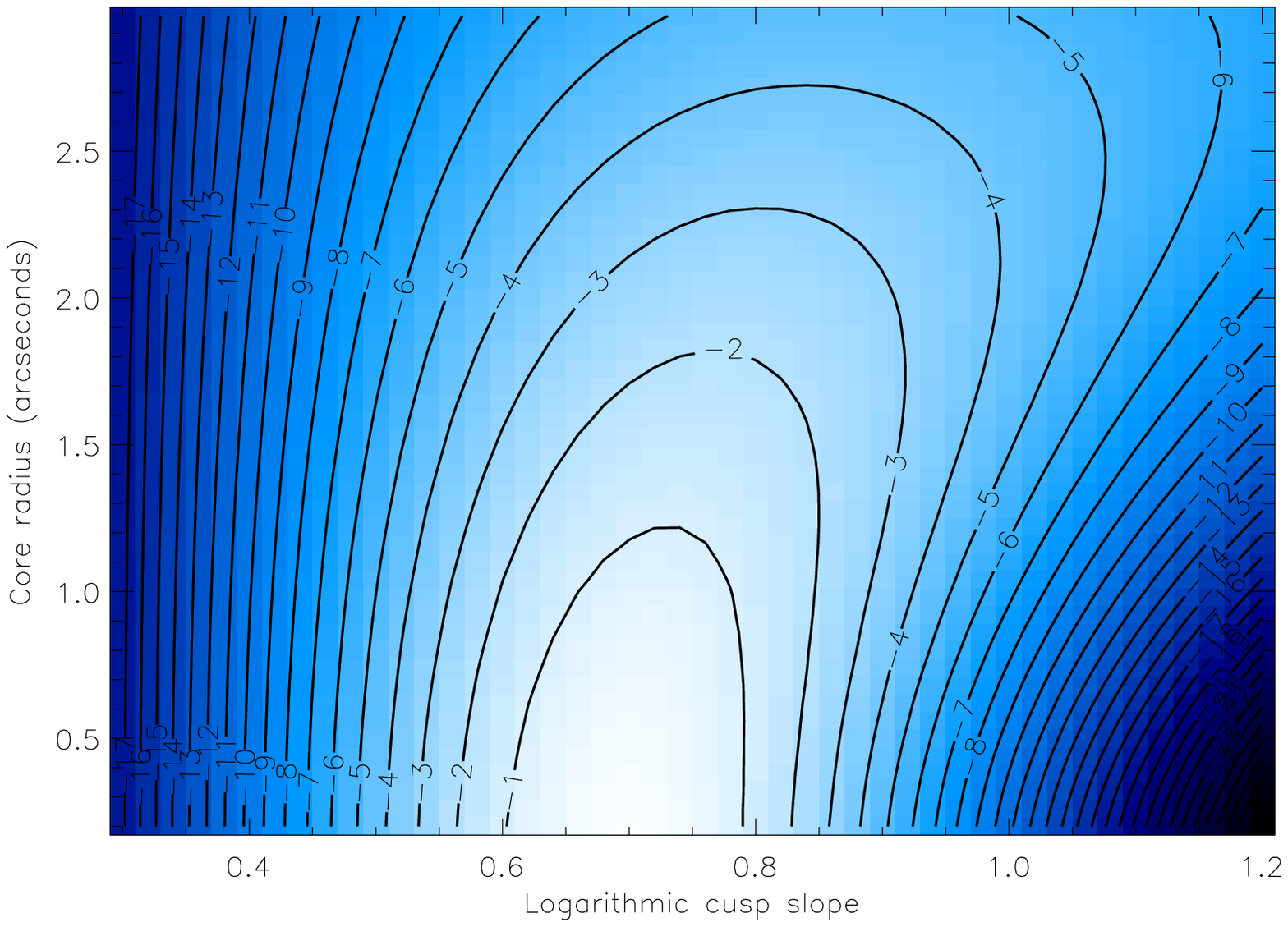}

Figure \ref{TurnoffLLF}
\end{figure}

\clearpage

\begin{figure}
\plotone{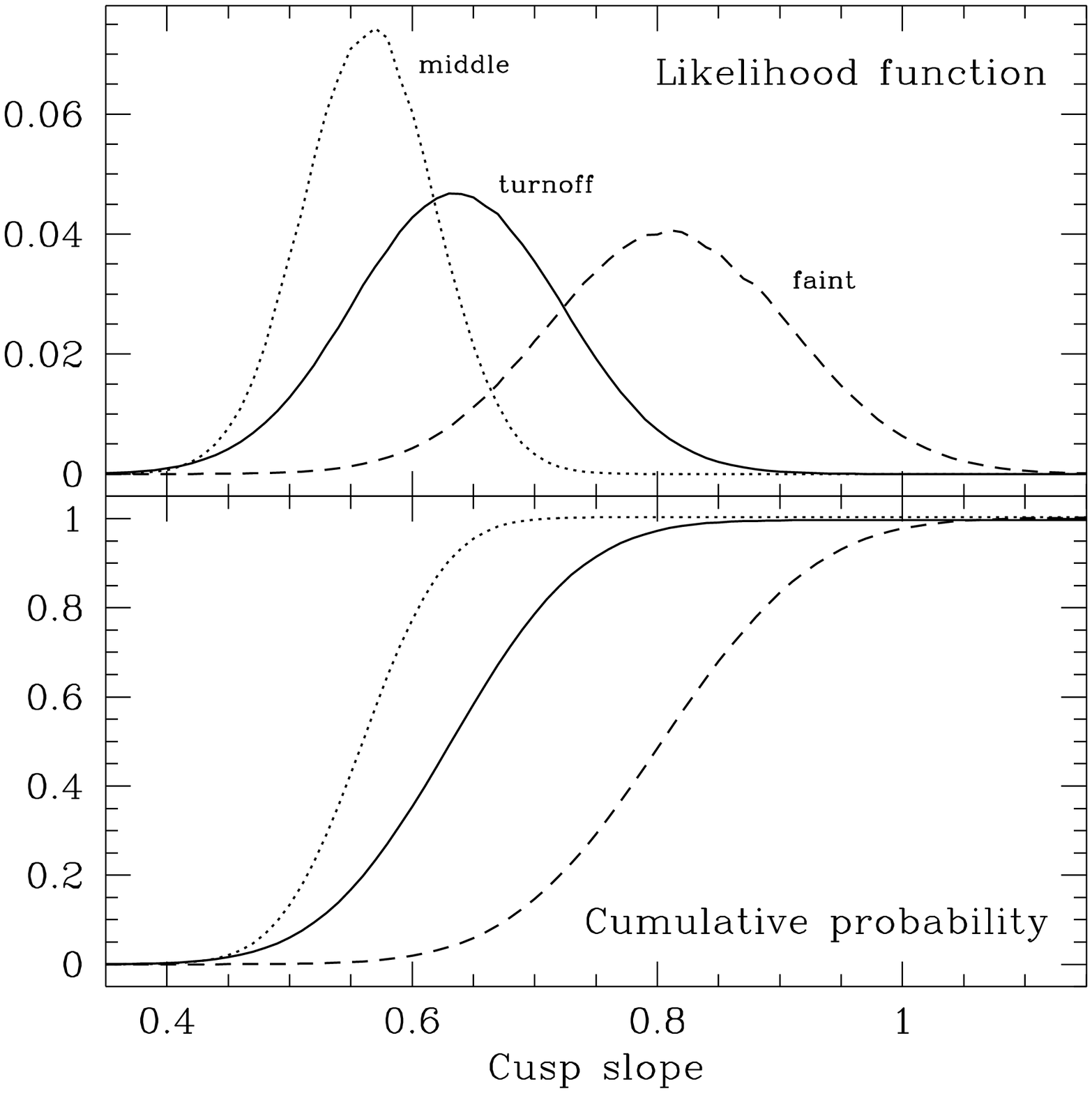}

Figure \ref{FaintLLF}
\end{figure}

\clearpage

\begin{figure}
\plotone{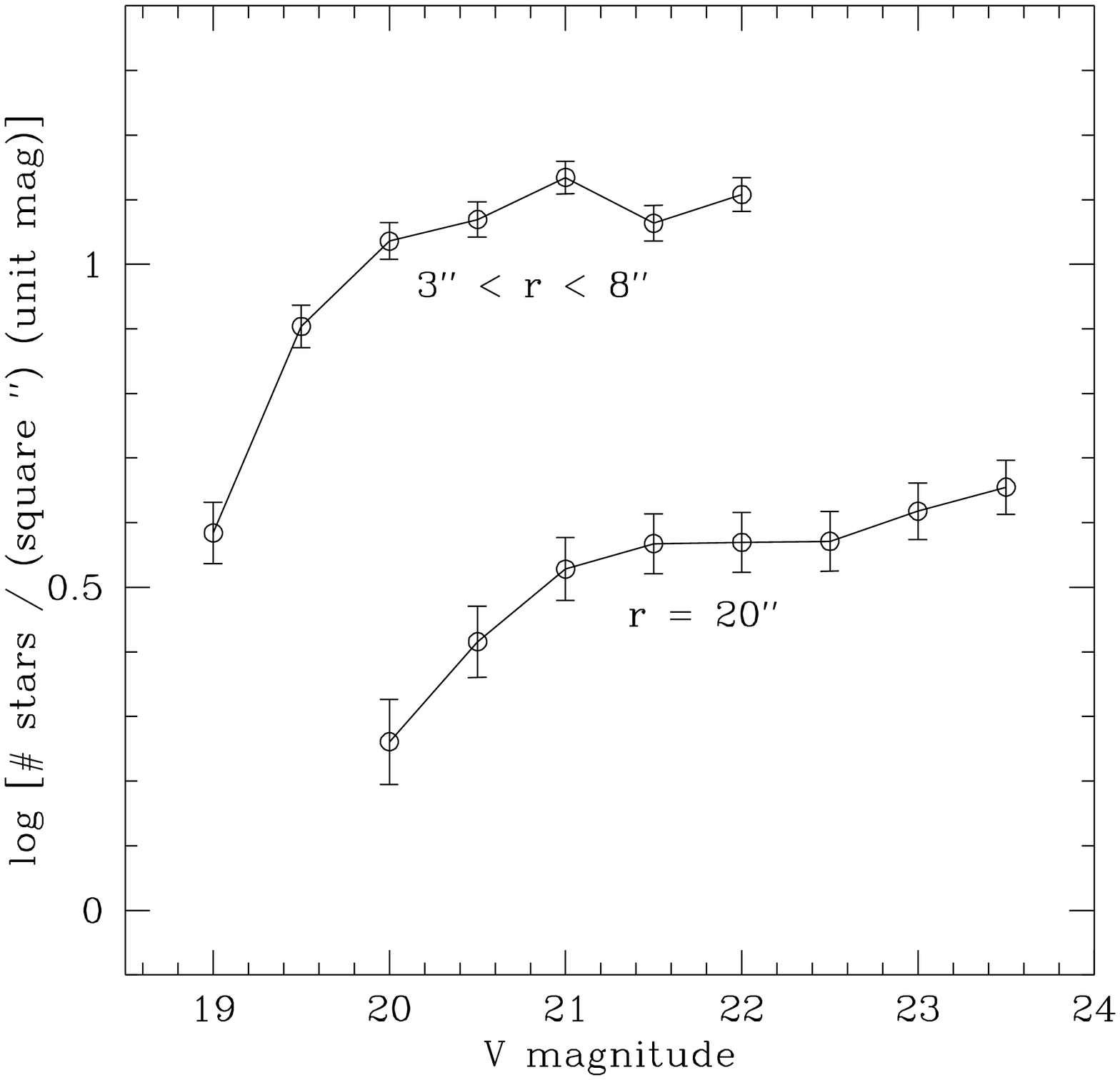}

Figure \ref{LumFunctions}
\end{figure}

\clearpage

\begin{figure}
\plotone{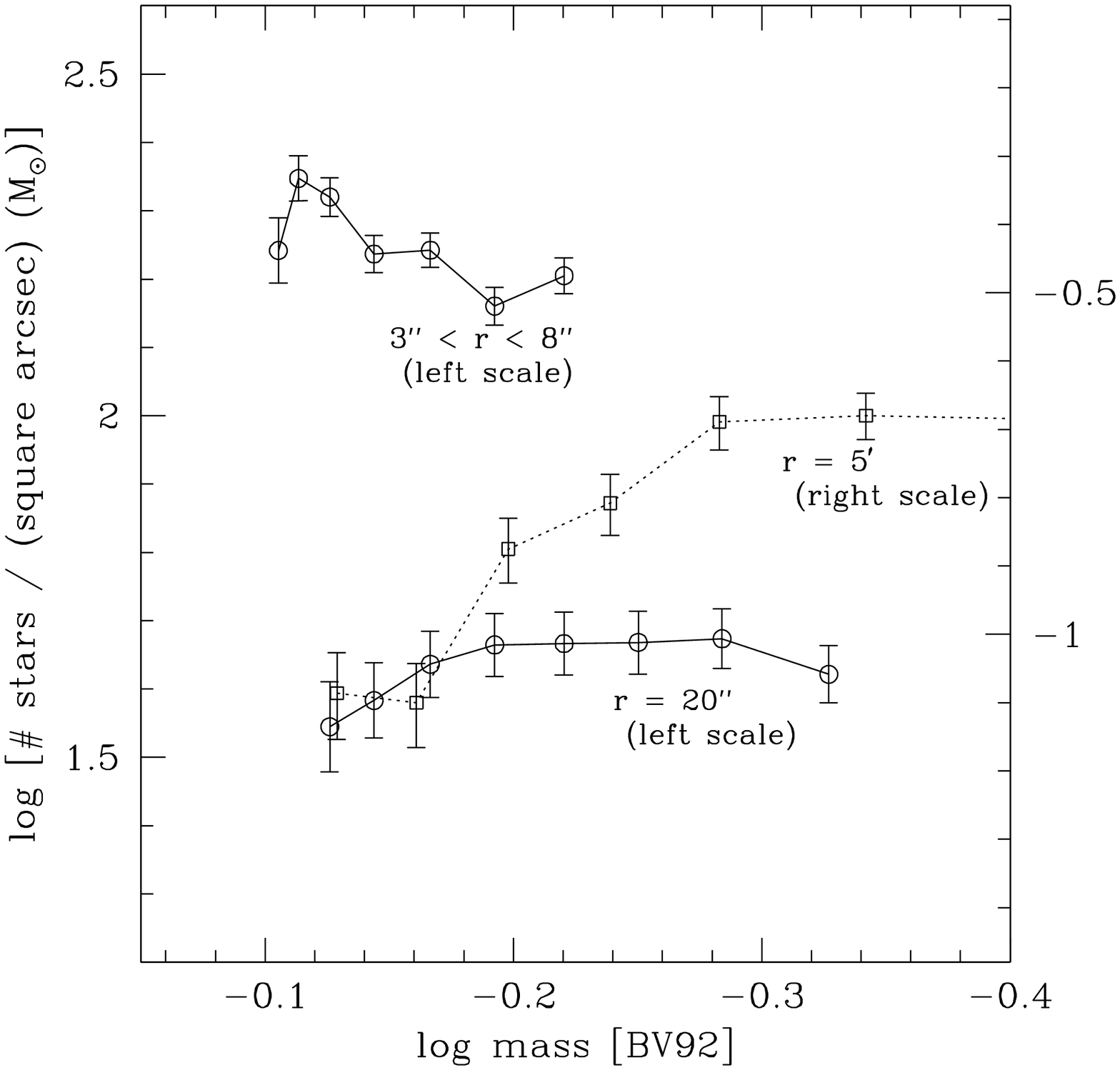}

Figure \ref{MassFunctions}
\end{figure}

\clearpage

\begin{figure}
\plotone{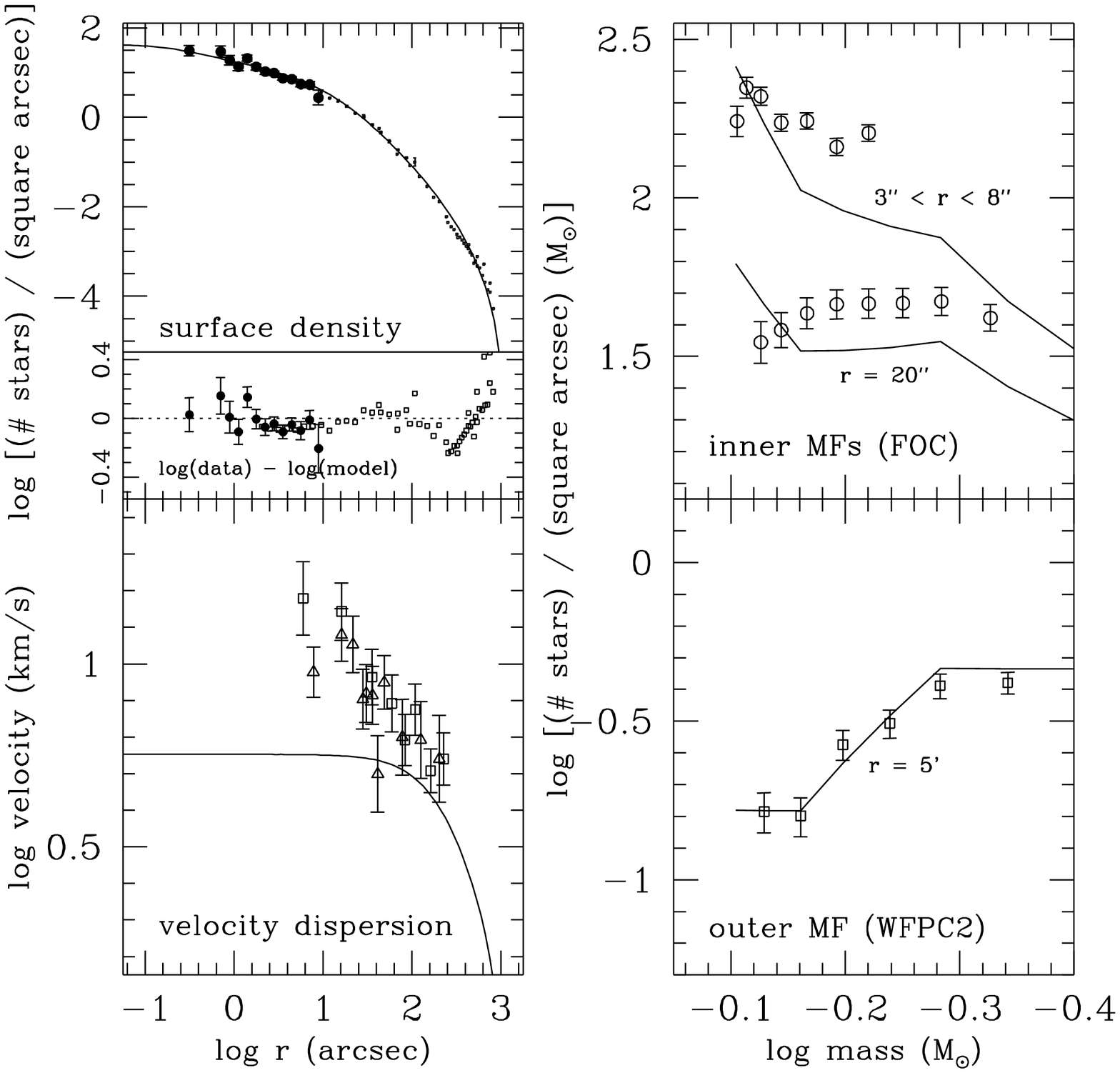}

Figure \ref{KingModel}
\end{figure}

\newpage

\begin{references}

\reference{bw76}
Bahcall, J.\ N., \& Wolf, R.\ A.\ 1976, \apj, 209, 214

\reference{bw77}
Bahcall, J.\ N., \& Wolf, R.\ A.\ 1977, \apj, 216, 883

\reference{BV92}
Bergbusch, P.\ A., \& VandenBerg, D.\ A.\ 1992, \apjs, 81, 163 (BV92)

\reference{bol89}
Bolte, M.\ 1989, \apj, 341, 168

\reference{cal93}
Calzetti, D., De Marchi, G., Paresce, F., \& Shara, M.\ 1993, \apj,
402, 1

\reference{cw90}
Chernoff, D.\ F., \& Weinberg, M.\ D.\ 1990, \apj, 351, 121

\reference{hc80}
Cohn, H.\ 1980, \apj, 242, 765

\reference{hc85}
Cohn, H.\ 1985, in Dynamics of Star Clusters, eds.\ J.\ Goodman \& P.\
Hut, p.\ 161

\reference{dmp96}
De Marchi, G., \& Paresce, F.\ 1996, ``Very Blue Stars and Mass
Segregation in the Core of M15'', preprint

\reference{djk84}
Djorgovski, S., \& King, I.\ R.\ 1984, 277, L49

\reference{dru88}
Drukier, G.\ A., Fahlman, G.\ G., Richer, H.\ B., \& VandenBerg, D.\
A.\ 1988, \aj, 95, 1415

\reference{dru95}
Drukier, G.\ A.\ 1995, \apjs, 100, 347

\reference{dmm94}
Dubath, P., Meylan, G., \& Mayor, M.\ 1994, \apj, 426, 192

\reference{DH93}
Durrell, P.\ R., \& Harris, W.\ E.\ 1993, \aj, 105, 1420 (DH93)

\reference{geb95}
Gebhardt, K., \& Fischer, P.\ 1995, \aj, 109, 209

\reference{geb94}
Gebhardt, K., Pryor, C., Williams, T.\ B., \& Hesser, J.\ E.\ 1994,
\aj, 107, 2067

\reference{gra92}
Grabhorn, R.\ P., Cohn, H.\ N., Lugger, P.\ M., \& Murphy, B.\ W.\
1992, \apj, 392, 86

\reference{gre91}
Greenfield, P.\ \etal\ 1991, \procspie, 1494, 16

\reference{GYSB}
Guhathakurta, P., Yanny, B., Schneider, D.\ P., \& Bahcall, J.\ N.\
1996, \aj, 111, 267 (GYSB)

\reference{gg79}
Gunn, J.\ E., \& Griffin, R.\ E.\ 1979, \aj, 84, 752

\reference{iau}
Hut, P., \& Makino, J.\ 1996, eds., ``Dynamical Evolution of Globular
Clusters:\ Confrontation of Theory and Observation'' (IAU Symp.\ 174)

\reference{jed94}
Jedrzejewski, R.\ I., Hartig, G., Jakobsen, P., Crocker, J. \ H., \&
Ford, H.\ C.\ 1994, \apjl, 435, 7L

\reference{irk65}
King, I.\ R.\ 1965, \aj, 70, 376

\reference{irk66}
King, I.\ R.\ 1966, \aj, 71, 64

\reference{irk68}
King, I.\ R., Hedemann, E., Hodge, S.\ M., \& White, R.\ E.\ 1968,
\aj, 73, 456

\reference{irk94}
King, I.\ R., Anderson, J., \& Sosin, C.\ 1994b, in Calibrating Hubble
Space Telescope:\ Proceedings of a Workshop Held at STScI, eds.\
J.C. Blades \& S.J.\ Osmer (Baltimore:\ STScI), p.\ 130

\reference{lau87}
Lauer, T.\ R.\ \etal, \apj, 369, L45

\reference{lug87}
Lugger, P.\ M., Cohn, H., Grindlay, J.\ E., Bailyn, C.\ D., \& Hertz,
P.\ 1987, \apj, 320, 482

\reference{mm91}
Meylan, G., \& Mayor, M.\ 1991, \aap, 250, 113

\reference{mic63}
Michie, R.\ W., \& Bodenheimer, P.\ 1963, \mnras, 126, 269

\reference{psc89}
Peterson, R.\ C., Seitzer, P., \& Cudworth, K.\ M.\ 1989, \apj, 347,
251

\reference{gp96}
Piotto, G., Cool, A.\ M., \& King, I.\ R.\ 1996, ``A Comparison of
Deep \hst\ Luminosity Functions'', submitted to \aj

\reference{nr}
Press, W.\ H., Teukolsky, S.\ A., Vetterling, W.\ T., \& Flannery, B.\
P.\ 1992, Numerical Recipes: The Art of Scientific Computing, Second
Edition (Cambridge: Cambridge University Press)

\reference{sar80}
Sarazin, C.\ 1980, \apj, 236, 75

\reference{me95}
Sosin, C., \& King, I.\ R.\ 1995, \aj, 109, 639

\reference{pbs87}
Stetson, P.\ 1987, \pasp, 99, 191

\reference{pbs92}
Stetson, P.\ 1992, in Astronomical Data Analysis Software, ed. D.\ M.\
Worrall, C.\ Biemesderfer, and J.\ Barnes (ASP Conference Series,
Vol.\ 25), p. 297

\reference{yan94}
Yanny, B., Guhathakurta, P., Bahcall, J.\ N.\, \& Schneider, D.\ P.\
1994, \aj, 107, 1745

\end{references}
\end{document}